%% file: main.tex
\documentclass{article} % For LaTeX2e
\usepackage[preprint]{neurips_2024}
\usepackage{microtype}

\usepackage{graphicx} % Required for inserting images
\usepackage{lineno}
\usepackage{url}
\usepackage{caption}
\usepackage{subcaption}
\usepackage{listings}
\usepackage{multirow}
\usepackage{booktabs}
\usepackage{wrapfig}
\usepackage{subfig}

\begin{document}

\title{LongFuncEval: Measuring the effectiveness of long context models for function calling}

\author{Kiran Kate \quad 
   Tejaswini Pedapati \quad
   Kinjal Basu \quad
   Yara Rizk \quad
  Vijil Chenthamarakshan \\
  \And
  \hspace{16mm} Subhajit Chaudhury \quad
  Mayank Agarwal \quad
  Ibrahim Abdelaziz \\
  \And
   \textbf{IBM Research}  \\
   \texttt{\{kakate, tejaswinip, ecvijil\}@us.ibm.com},  \\
   \texttt{\{Kinjal.Basu, yara.rizk, subhajit, mayank.agarwal, ibrahim.abdelaziz1\}@ibm.com}}

\maketitle

\begin{abstract}
\input{sections/abstract}
\end{abstract}

\section{Introduction}
\input{sections/introduction}

\section{Related Work}
\input{sections/related_work}

\section{Problem Definition}
\input{sections/problem_definitions}

\section{Methodology and Experimental Setup}
\input{sections/methodology}

\input{sections/experimental_setup}
\section{Results}
\input{sections/results}

\section{Conclusion}
\input{sections/conclusion}

\section*{Ethics}
For the study we performed on long context and function calling, we do not foresee any ethical concerns or opportunities to misuse this work. Furthermore, the data we used (and modified) to evaluate models does not contain any PII or harmful language. 

\bibliographystyle{aaai}
\bibliography{main}

\newpage
\appendix
\section*{Appendix}
\input{sections/appendix}
\end{document}

%% file: sections/abstract.tex
Multiple recent studies have documented large language models' (LLMs) performance on calling external tools/functions. Others focused on LLMs' abilities to handle longer context lengths. At the intersection of these areas lies another interesting problem: LLMs' abilities to accurately perform function calls in long context settings. Particularly, when calling tools, LLMs are encumbered by three predominant challenges: (1) a large catalog of tools, (2) long responses from the tool APIs, and (3) long multi-turn conversations. These challenges are particularly relevant to enterprise applications of LLMs which engage in multi-turn conversations with users to complete complex tasks that require a large catalog of complex tools. The literature contains multiple investigations of long context challenges such as lost in the middle or needle in the haystack for natural language tasks. In this paper, we make the first attempt to comprehensively study the long context understanding capabilities of these models in the tool calling setup. We modify existing benchmarks for challenge 1 and 3, and create a new evaluation set for challenge 2 to enable this analysis. We gradually increase the input context length and also vary the position of the answer in the input.
% Our empirical analysis across 7 model size reveals significant performance degradation as the number of tools grows and as tool responses become larger
When evaluated with several long context models, we observe a performance drop of 7\% to 85\% as the number of tools increases, a 7\% to 91\% degradation in answer retrieval as the tool responses length increases, and 13\% and 40\% degradation for as  multi-turn conversations get longer. Our study shows that LLMs still struggle with long context in tool calling settings, motivating future research to drive further LLM improvements.  

%% file: sections/introduction.tex
Recent advances in large language models (LLMs) have led to exciting use cases in several domains such as coding, education, travel, and healthcare. However, they still face key challenges such as struggling with mathematical reasoning, accessing external knowledge, not having access to sensitive data for training (e.g., Personally Identifiable Information), and not having access to most up-to-date knowledge. To address the aforementioned limitations, researchers have focused on teaching LLMs how to utilize external resources via tool calling\footnote{We use the terms tool, function, and API interchangeably.}~\citep{li2023api,qin2024toolllm}, rather than overwhelming models with additional data during training or fine-tuning stages. This approach allows LLMs to tap into sensitive external data, use tools, reasoning systems, and code interpreters, and most importantly, retrieve up-to-date information — an essential capability for real-world applications.

% To tackle real-world task automation problems, large language models (LLMs) must augment their parametric knowledge with external tools. This ability, called function/tool calling\footnote{We use the terms function, tool, and API interchangeably.}, has been the focus of many recent works \citep{li2023api, qin2024toolllm}. %, schick2023toolformer,  patil2024gorilla, , wang2024tools, qu2025tool}. 
% Crucial in real-world applications, this enables access to up-to-date information, accurate computations, data retrieval, and integration with other systems for more reliable and dynamic responses.

In different use cases, besides tool calling, LLMs are used to solve complex tasks such as carrying on long conversations \citep{Maharana2024_long_conv}, answering questions from documents \citep{li2023loogle, liu2024longgenbench}, reasoning using facts spread across several large documents \citep{kuratov2024babilong}, etc. 
This requires LLMs to have long context comprehension capabilities and thus heralded the era of long context models. However, studies \citep{niah,liu-etal-2024-lost,hsieh2024ruler} have shown that LLMs exhibit position bias and cannot fully utilize the entire information provided to them. This bias exacerbates as the input context length increases. Specifically, \citep{niah} observed that LLMs' performance deteriorates when the relevant information is present in the middle of a large context and dubbed it ``lost-in-the-middle''. Long context understanding of LLMs has been evaluated on various natural language tasks such as needle-in-a-haystack~\citep{niah, hsieh2024ruler}, passkey retrieval~\citep{mohtashami2023randomaccess}, summarization~\citep{shaham2023zeroscrolls}, and question answering~\citep{hsieh2024ruler}. Many of these tasks are simple and synthetic and do not cover any of the tool-calling aspects of LLMs.
% To the best of our knowledge, we are the first paper to evaluate LLMs in a long context tool calling setup.

 % Multiple research studies have documented varying flavors of the long context challenge \citep{niah,liu-etal-2024-lost, hsieh2024ruler}, and others developed schemes to extend models' context \citep{peng2023yarn, han2023lm}. %\citep{gao2024train, chen2023longlora}.%, jin2024llm, lu2024controlled}. 
% \textcolor{blue}{Tejaswini, should we replace these two citations with more popular Yarn, LM-infinite?}

\begin{figure}[t]
    \centering
    \includegraphics[width=1\textwidth]{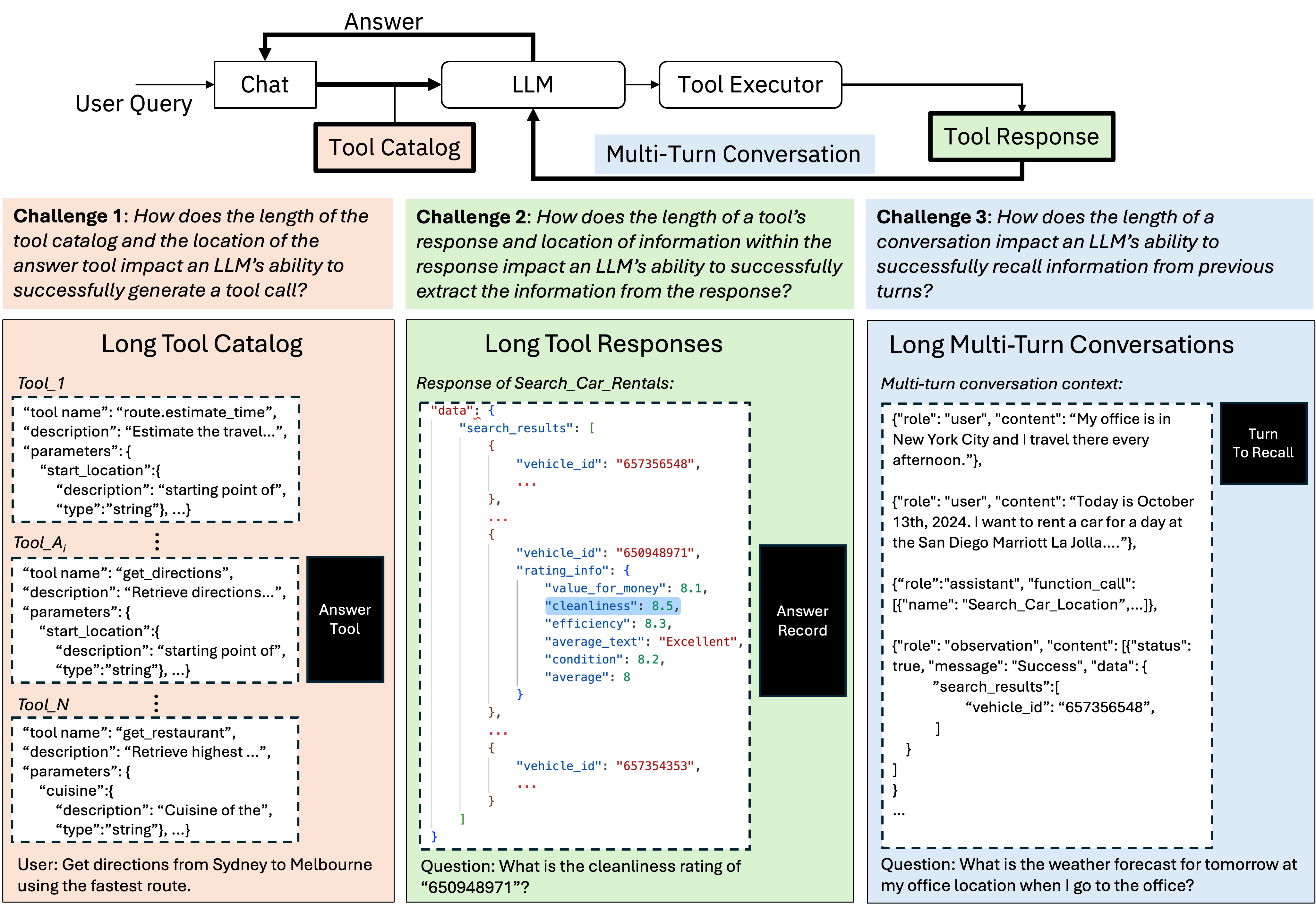}
    \caption{Overview of the three challenges at the intersection of long context and tool calling}
    \label{fig:overview}
\end{figure}

Recognizing tool usage and long-context handling as essential capabilities of LLMs to build real-world applications, in this paper, we investigate LLMs' performance in a long-context tool-calling setup. As tool calling evolved from using simple tools such as a calculator and a calendar to more complex scenarios, LLMs face multiple manifestations of the long context challenge that are specific to tool calling. The first relates to the \textbf{large number of tools} in the catalog (hundreds of tools for real-world setups). %; some having hundreds of arguments and wordy descriptions, preventing all this information from fitting in an LLM's context. 
The second relates to the \textbf{long tool responses}; real-world tools are often wrapped in Rest APIs, which have large JSON responses that in extreme cases cannot fit in 128K context length. 
% The default practice in multi-turn tool calling is to simply append tool responses to the context, and depend on LLMs' ability to extract information from them to plan the next step.
The third relates to \textbf{dependent turns in multi-turn dialogues}, where information from one user turn or tool response is used in future turns, and long multi-turn conversations make it hard for the models to understand and retrieve relevant parts of the context to generate the next function call. Figure \ref{fig:overview} illustrates these three challenges. 

In the tool calling literature, the long tool catalog challenge is mainly handled with independent tool retrievers~\citep{lumer2024toolshed, qin2024toolllm}, long responses are handled by truncating or parsing them~\citep{qin2024toolllm, song2023restgpt}, and long multi-turn conversation settings are evaluated by existing benchmarks~\citep{berkeley-function-calling-leaderboard, zhong2025complexfuncbench}. However, none of the existing works study the behavior of LLMs on these tasks as the size of the context becomes larger. Additionally, there are no insights on whether the location of information in the context affects model performance for tool calling.

In this paper, we evaluate various SOTA LLMs that are popular for tool-calling and also support long contexts. Despite having 128K context length, we observe varying degrees of degradation across model families and model sizes. When varying the tool catalog size (from 8K to 120K tokens) and location of the correct tool, performance drops vary from 7.59\% to 85.58\% (excluding Mistral-large) across all the datasets, due to formatting issues and increased hallucination. When varying the length of the response (from 10K to 80K tokens), degradation in answer retrieval varies from 7\% (GPT-4o) to about 91\% (Mistral-Large). Finally, when varying the length of the conversation (from 10K to 80K tokens), between 13\% and 40\% degradation was observed. 

Our contributions can be summarized as follows.
(1) To the best of our knowledge, we are the \textbf{first paper to comprehensively study the challenges in long context tool calling}. We identify three key challenges and design evaluation setups by extending public benchmarks to study LLM behavior for each challenge. 
(2) We create a \textbf{new evaluation dataset for challenge 2} to evaluate LLM's ability to handle long tool responses. 
(3) We evaluate \textbf{six 128K context length tool calling open-weight models and one closed-weight model} and show that LLMs struggle with all three challenges.

%% file: sections/related_work.tex
\subsection{Tool Calling}
% Define terminology: tools vs. functions vs. APIs

% A vast number of papers have been published on tool-calling models \citep{abdelaziz2024granite, patil2024gorilla, lin2024hammer}, %, liu2024toolace, srinivasan2023nexusraven, hao2023toolkengpt, chen2024octopus}, 
% and training and evaluation benchmarks to drive progress in the field \citep{qin2024toolllm, basu2024nestful, liu2024apigen}%, han2025nestools, wu2024seal, guo2024stabletoolbench, basu2024api, guo2024api, jin2024toolbridge, huang2023metatool, li2023api}
% . Various techniques have been used to improve tool calling performance such as 1) creating good quality synthetic datasets based on verifiable API traces \citep{liu2024apigen, yin2025magnetmultiturntoolusedata,liu2024toolace}, 2) utilizing model alignment techniques  \citep{yin2025magnetmultiturntoolusedata, yu2025steptoolenhancingmultisteptool, zeng2025itoolboostingtooluse}, 3) formulating tool calling as a multi-task problem \citep{abdelaziz2024granite}, %4) extending models' context window \citep{zhong2025complexfuncbench}, 5) employing prompt optimization techniques \citep{lumer2024toolshed}, 
% and many others. 

% A related research problem is understanding how tool-calling models perform when presented with a large number of tools.
Tool calling involves generating one or more function calls in response to a user query. A sub-task of tool calling is tool selection from a catalog. 
% This issue is closely related to Challenge 1 in our study. 
While many papers have explored the tool selection problem, acknowledging the challenges related to catalog size, we did not find any studies that systematically documented the impact of tool catalog size on tool calling performance of LLMs. %For instance, TinyAgent \citep{erdogan2024tinyagent} found that prompting an LLM with all 16 available tools resulted in 78.89\% accuracy, while using a tool retriever to shortlist just three tools slightly improved accuracy to 80.06\%. Although limited in scale, this finding suggests that increasing the number of available tools may negatively impact model performance. Similarly, \citep{lumer2024toolshed} demonstrated a decline in a tool retriever's performance as the number of input tools increased from 1 to 4,000, further highlighting the challenges LLMs face when dealing with large tool catalogs and longer contexts in general.
Some used small catalog sizes 
%that fit within a model's context window 
\citep{erdogan2024tinyagent}, others relied on a tool retriever to shortlist relevant tools before selection \citep{patil2024gorilla, qin2024toolllm}%, gao2024confucius, xu2023tool, song2023restgpt, kong2023tptu, yuan2024craft, liu2024toolnet, du2024anytool} 
, even proposing novel retrieval approaches \citep{xu2024enhancing, shi2025retrieval}. \citep{lumer2024toolshed} demonstrated a decline in a tool retriever's performance as the number of input tools increased, but did not document impact on tool calling when all the tools are used in the tool calling prompt. %we did not find any studies that systematically documented the impact of tool catalog size on models' tool calling performance. 

% Beyond large catalogs, additional factors contribute to longer context in tool-calling models. The first is tool response; the literature typically appends results of tool executions to a prompt \citep{zhong2025complexfuncbench}. These responses can vary significantly in length; some tools return concise outputs (e.g., a single number in MathQA  \citep{basu2024nestful}). Others generate long, complex JSON responses with extensive metadata~\citep{qin2024toolllm}. This challenge is similar to NL2SQL tasks, where large tables are retrieved \citep{liu2024survey, chung2025long}, increasing processing complexity and impacting performance \citep{nguyen2025sqlong}.

While most tool calling benchmarks include simpler tools producing simple responses, ToolLLM~\citep{qin2024toolllm} proposes to handle long tool responses by asking GPT-4 to drop irrelevant keys and/or truncating it beyond a certain number of tokens. Another approach to handle long responses from REST APIs is to parse them using Python code~\citep{song2023restgpt}. Recent benchmarks such as BFCL-v3\footnote{\url{https://gorilla.cs.berkeley.edu/blogs/13_bfcl_v3_multi_turn.html}} specifically evaluate model performance in multi-turn long context scenarios. Similarly, ComplexFuncBench \citep{zhong2025complexfuncbench} is an evaluation dataset handling complex tools, long parameter values, and long multi-turn tool calling scenarios.

% In some BFCL categories, for instance, models must execute tools sequentially, using each output as input for the next call. Failures in tool execution further complicate matters, as error messages accumulate in the conversation history, sometimes leading to input prompts that quickly exceed the context length. This category is particularly challenging, with even top-tier models like GPT-4o achieving below 50\% accuracy.
Unlike these works, our goal is to study whether LLMs can extract required information from API responses or recall information from previous conversation turns, especially as the context becomes larger. To the best of our knowledge, we did not find any systematic studies that explore the three long context tool calling challenges.

\subsection{Long Context}
% How these don't tackle the function calling specific challenges: 

% To cater to the growing demand for tasks entailing increased context lengths \citep{liu2025comprehensive}, recent models such as Gemini \citep{Anil2023_Gemini}, Claude, GPT-4 \citep{OpenAI_GPT4}, LLama 3.1 \citep{grattafiori2024llama}, DeepSeek \citep{bi2024deepseek}, Granite \citep{stallone2024scaling} 
% % and others extended their context 
% have a context length of at least 128K tokens. 
Numerous datasets benchmark models in long context settings including Zero-Scrolls \citep{shaham2023zeroscrolls}, Longgenbench \citep{liu2024longgenbench}, Loogle \citep{li2023loogle}, NoLiMa \citep{modarressi2025nolima}, DENIAHL \citep{dai2024deniahl}.
% , FACTOR \citep{liufactor}, HELMET \citep{yenhelmet}, 
% and ComplexFuncBench \citep{zhong2025complexfuncbench}. However, studies still show that models are struggling with long context.  

LLMs struggle with long-context comprehension, particularly when key information is buried within large inputs. The ``Needle-in-a-Haystack`` study showed that LLMs struggled to find key information when presented with content with large context \citep{niah, das2024needle}. Similarly, the ``Lost in the Middle`` study found that model performance declines when the critical information is positioned at different points within the prompt, highlighting challenges in attention mechanisms \citep{liu-etal-2024-lost}. RULER studied the effective context length for models claiming to have at least 32K context length and found the actual context length was not always as claimed \citep{hsieh2024ruler}. 
These challenges arise primarily due to long term decay property of RoPE position embedding \citep{Su2024_Rope} that is widely adopted in most of the recent transformers 
and dispersion (difficulty) and scope (amount) of information retrieval tasks \citep{goldman2024really}. While most of these studies focus on unstructured text, our work specifically investigates long-context challenges in structured tool-calling scenarios, where models must process and retrieve relevant information from complex, structured outputs.

% Method to extend context
% Passkey retrieval~\citep{mohtashami2023randomaccess}
% U-NIAH \citep{gao2025u}

%% file: sections/problem_definitions.tex
To concretely define the long context tool calling challenges, consider a travel planning agent which has to find available dates for all the participants, book the flights, hotels, rental cars,  reserve tables at restaurants and recommend activities to do based on the group's interest. This agent's pipeline consists of the steps defined in Figure \ref{fig:overview}. The agent must first access the group's calendars and find suitable dates for their trip, then book flights, hotels, reserve restaurant tables, etc., interacting with the users at various steps to obtain their preferences. Towards successfully completing this task with real world APIs, an LLM is faced with the following three challenges: 

\subsection{Challenge 1: Large Tool Catalog}

At each step, the travel planner must access a large catalog of tools. This catalog consists of tool names, their descriptions, and their input-output arguments. Depending on how rich and complex these tools are, a 128K token context window may not even fit all tools (see Table \ref{tab:num_tools} in the Appendix); as seen in multiple benchmarks, catalog sizes can easily reach hundreds if not thousands of tools \citep{liu2024apigen}. 

Formally, let \( \mathcal{P} \) denote the input prompt, which consists of a user query \( u \) and a list of available tools \( \mathcal{T} = \{t_1, t_2, \dots, t_n\} \). The user query \( u \) is a natural language expression of a task or problem that needs to be solved. The tool list \( \mathcal{T} \) contains descriptions including parameter names, descriptions and types, each of which is represented as \( t_i = (name_i, description_i) \), where \( name_i \) is the unique identifier of the tool and \( description_i \) is the detailed functionality of the tool. Given \( \mathcal{P} = (u, p, \mathcal{T}) \), the LLM generates a tool call \( f \) from the tool list \( \mathcal{T} \) that satisfies the user query \( u \). Here, \( p \) is the position of tool/s in \( \mathcal{T} \) which correspond to the ground truth function call/s.

% The LLM performs this by semantically matching the user query \( u\) with the descriptions \( description_i \) of the tools in \( \mathcal{T} \), and selecting the most appropriate tool \( t_i \) to generate the function \( f \).

Our goal is to study how the size of the catalog \( n \) impacts the ability of an LLM (that can handle large contexts) to formulate the correct tool call. We also study how the location of the answer tool/s within the context impacts the model's performance. 

% Our formulation of the large tool catalog problem assumes a typical tool calling setting where LLMs are prompted with a natural language utterance from the user ($U_{NL}$), a list of $N$ tools with their names and specifications such as descriptions, parameter names, descriptions, types etc. ([$tool\_1, ..., tool\_A_{i}, tool\_A_{i+1}, ..., tool\_A_{i+j}, ..., tool\_N$]). The task for the LLM is to output one or more function calls to address the user query. In this case, the tools at index $A_{i}$ to $A_{i+j}$ correspond to the correct tools to be called. 
% To evaluate the model's efficacy in handling a vast amount of tools, we gradually increase $N$ and compute tool calling performance at various values of $N$. Furthermore, similar to~\citep{liu-etal-2024-lost, niah, mohtashami2023randomaccess}, we also vary the position of the answer, i.e $A_{i}$ to study if the position of the answer has an impact on the tool calling performance.

\subsection{Challenge 2: Long Tool Responses}
Once LLMs formulate a tool call (by selecting a name from a catalog of tools and determining the arguments' values), the tool executor executes the tool. It is common practice to simply append the tool response to the prompt and send it back to the LLM to process the response and plan the next steps. For many real world tool calling scenarios, the tools are REST APIs~\citep{qin2024toolllm, zhong2025complexfuncbench, ruan2024toolemu} which return responses in JSON format. REST API responses are long and highly nested and existing approaches either compress or truncate the responses beyond a token limit~\citep{qin2024toolllm}. 

A REST API response typically returns an array of information, say a list of hotels, or flight paths. Each record in the array is usually a nested JSON object. In \emph{tool response QA}, given the triplet \( (\mathcal{Q},\  p,\  \{r_1, r_2, ..., r_n\}) \), where \( \{r_1, r_2, ..., r_n\} \) is a list of records in the tool response, the task is to prompt the LLM to generate an answer to the question \( \mathcal{Q} \) based on the tool response. The index \( p \) corresponds to the location of a record within tool response which has the information to answer \( \mathcal{Q} \).

Our goal is to study how well LLMs understand tool responses and whether they can retrieve correct information from them, especially when the responses are large and complex. To the best of our knowledge, no existing tool calling datasets enable evaluation of this aspect. We propose a new task called \emph{tool response QA}, which creates question-answer (QA) pairs on tool responses from an existing benchmark~\citep{zhong2025complexfuncbench}\footnote{We will make this dataset available to facilitate further research on tool response understanding.}.

\subsection{Challenge 3: Long Multi-Turn Conversations}
% Existing tool calling benchmarks identify multi-turn/multi-step tool calling as an important setting and show that even the best tool calling LLMs find it challenging ~\citep{berkeley-function-calling-leaderboard, wang2024mint, basu2024nestful}. The high number of turns or steps leading to long context is also acknowledged~\citep{berkeley-function-calling-leaderboard, zhong2025complexfuncbench} but none of the previous evaluation focused on systematic evaluation of the number of turns or number of tokens leading to drop in multi-turn tool calling. We focus on that aspect in challenge 3.

The travel planning task cannot be accomplished in a single conversational turn; in fact it requires multiple turns. 
As multi-turn tool-use benchmarks become more prevalent, it is evident that even the best-performing models struggle in these settings ~\citep{berkeley-function-calling-leaderboard, wang2024mint, zhong2025complexfuncbench}. A key challenge that remains largely unaddressed is the ability to effectively handle scattered information across multiple turns, essentially managing long-context dependencies and retrieving relevant information when needed. 
%For instance, a user might ask for flight details at the beginning of a conversation, and a flight-booking tool provides the relevant information. Later, the user may request a hotel near their arrival airport; if the model fails to retain and retrieve the earlier flight details, it will struggle to generate an accurate query for the hotel-booking tool. This highlights the difficulty of managing long-context dependencies and retrieving relevant information when needed, making multi-turn tool use a complex and unsolved problem. In this paper, we study the effect of this problem on multiple state-of-the-art models.

Specifically, we evaluate the long dependency problem~\citep{yin2025magnetmultiturntoolusedata} for multi-turn tool calling. The problem formulation has a set of user, LLM, and tool turns \( \{ u_1,\  llm_1,\  tool_1, \ ...,\  u_i,\  ...,\  u_n,\  llm_n \} \) where either \( u_i \) or \( tool_i \) contains information that is required to generate the function call \( llm_n \) on the user turn \( u_n \). We increase the number of the turns \( n \) in these conversations and study if it affects the ability of the models to recall the information from the \( i^{th} \) turn to generate a correct function call.

% We study the effect of this problem on multiple state-of-the-art models. Specifically, we investigate the impact of the number of turns (between a piece of information and a tool call that needs it) impacts an LLM's ability to successfully execute the tool call. 

%% file: sections/methodology.tex
\subsection{Challenge 1: Large Tool Catalog}

% \subsubsection{Unique Function Extraction and Long Context Generation}

To create large tool catalogs suitable for our study, the first step involves extracting all the unique functions from the BFCL dataset~\citep{berkeley-function-calling-leaderboard}. Let \( \mathcal{D} \) denote the dataset containing multiple sources of tools (such as \texttt{simple}, \texttt{multiple}, \texttt{parallel}, \texttt{live\_simple}); each source containing a list of function descriptions. The extraction process identifies all unique functions \( \mathcal{T}^{*} = \{t_1, t_2, \dots, t_N\} \) from \( \mathcal{D} \), eliminating duplicates. This is achieved by checking for uniqueness based on the function name.

Once the unique functions \( \mathcal{T}^{*} \) are obtained, the next step is to generate long context scenarios by introducing distractors. Distractors are functions that are not relevant to the user query \( u \) but are randomly added to the context to simulate a large tool catalog. The distractors are selected from \( \mathcal{T}^{*} \) and inserted into the tool list. The original function is then inserted at a specified position \( p \) of the distractor list, which is determined by a fraction \( \alpha \) of the length of the distractor list to the get the final tool list. This process is repeated until the token limit is reached. This insertion process ensures that the context length is increased while maintaining the original order of the functions.

This method was applied for various context lengths to systematically study the performance of the LLM. Specifically, we considered the following context lengths: \( \{8,192 ; 16,384 ; 32,768 ; 65,536 ; 120,000\} \) tokens\footnote{Although several models have a max context length of 131,072, we used 120,000 to account for the maximum number of additional tokens introduced by tokenization and the prompts introduced by all the models. }. Additionally, similar to ~\citep{liu-etal-2024-lost, niah, mohtashami2023randomaccess}, we also choose  different insertion positions of the original tools in the tool list. This is controlled by an insertion fraction \( \alpha \) which took one of the following values: \(0.1, 0.3, 0.5, 0.7, 0.9\),
% \( \{0, 0.1, 0.2, 0.3, 0.4, 0.5, 0.6, 0.7, 0.8, 0.9, 1.0\} \) \textcolor{blue}{TP: We only ended up using 0.1, 0.3, 0.5, 0.7, 0.9}
to explore the impact of different insertion positions on the long context generation. This comprehensive variation in context lengths and insertion positions allows for a detailed analysis of the LLM's performance under different long context scenarios.

We used the \texttt{simple}, \texttt{live\_simple}, and \texttt{multiple} subsets from BFCL~\citep{berkeley-function-calling-leaderboard} for this evaluation which have 858 samples in total.  We evaluated 6 models with context 128K, Mistral-large, ToolACE-8B, BitAgent-8B, Granite-3.1-8b-instruct, llama-3-1-70b-instruct across the three datasets for all the positions. Owing to budget constraints, we ran GPT-4o-2024-11-20 for context lengths 8,192, 32,768 and 120,000 for positions 0.1, 0.5, and 0.9. Appendix \ref{appendix:models} includes the exact details of the models. We used BFCL's code \footnote{\url{https://github.com/ShishirPatil/gorilla/}} for the prompts and the evaluations and used the AST accuracy metric.

\subsection{Challenge 2: Long Tool Responses}
\label{2:methodology}
% Creation of the dataset 
To evaluate an LLM's ability to process tool responses, we created a QA dataset based on the API responses. 
%To create a QA dataset based on API responses,
We use tool responses from ComplexFuncBench~\citep{zhong2025complexfuncbench}, a dataset that enables evaluation of complex function calling scenarios with multi-turn conversations and REST API tools. Our questions are based on five REST endpoints from booking.com\footnote{\url{https://rapidapi.com/DataCrawler/api/booking-com15}} spanning domains such as flight reservation, hotel booking, car rental, and tourist attractions. 

\begin{figure}
    \centering
    \includegraphics[width=1\linewidth]{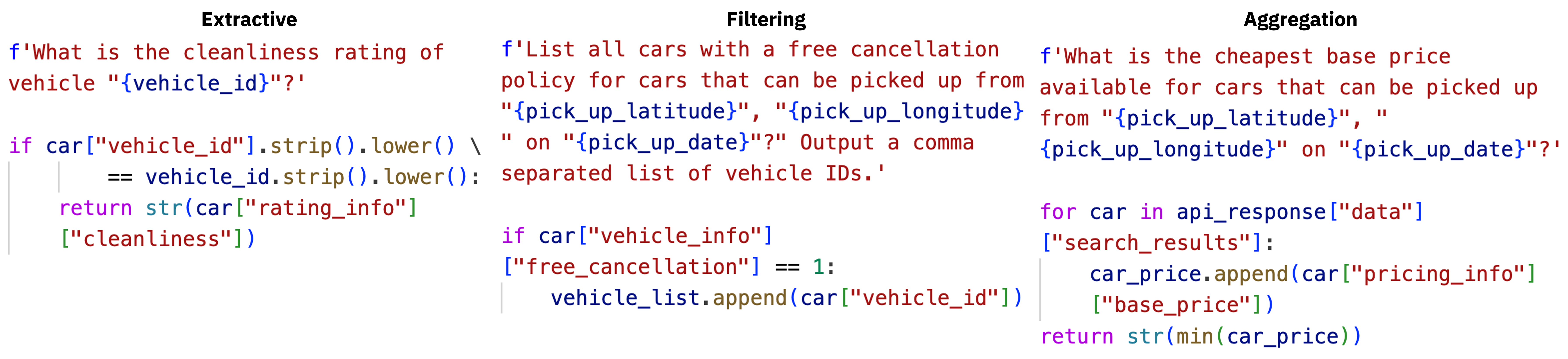}
    \caption{Examples of tool response question categories for challenge 2}
    \label{fig:json-QAType}
\end{figure}

Next, we manually create question templates for \( \mathcal{Q} \) and write code to parse the JSON to get answers to those questions by understanding the structure of these responses. The questions target three categories of response processing/understanding (as shown in Figure \ref{fig:json-QAType}): (1) extraction: extracting a single value for a key, (2) filtering: retrieving one or more entries based on some filtering criteria, or (3) aggregation: aggregating multiple entries by performing an aggregation operation to obtain the final answer (e.g., average, sum, etc.). %Figure \ref{fig:json-QAType} shows an example of each category.

The creation of such QA pairs is controlled by \( p \) which corresponds to the positional index of a record in the JSON corresponding to that QA pair. For example, for the extraction question in Figure \ref{fig:json-QAType}, the response contains information about different vehicles, and the value of \( p \) would decide what value gets substituted for $"vehicle\_id"$ in the question template. This control allows us to vary the position of the QA pair and evaluate if the position matters similar to other such studies ~\citep{liu-etal-2024-lost, niah}.

To study the impact of the size of tool responses, we control the number of records \( n \) such that it reaches a token limit. This is done by merging responses of individual API calls in the dataset which simulates changing the original API call to have fewer filtering arguments. 
For each API endpoint, we create 10 response samples such that the response with token limit \( n_1 \) is a prefix of response with token limit \( n_2 \) when \( n_1 \ < n_2 \).
Appendix \ref{appendix:data_details} includes details of the dataset created (which had a total of 566 samples).

% Experimental setup
We used zero-shot prompting with greedy decoding to get LLMs to generate the answers. For each question template, the evaluation metric depends on the type of ground truth. For example, a scalar value warrants an exact match but an unordered list would need to be matched discounting the order. From our experiments, we noticed that LLMs were very verbose in their responses even when prompted to output just the precise information. We added special instructions (e.g., "show only the amount and no currency or currency symbol.") at the end of the question to account for this. In addition, we also report results for a "contains" metric which checks for membership of the ground truth in the generated answer in appendix \ref{appendix:rq2_detailed_results}. 

We conduct two experiments: one by varying the tool response token limits to \( \{ 10K, 20K, 40K, 80K\}\) and by fixing \( p \) = 1, and second by fixing the tool response token limit to \( 80K \) tokens and varying \( p \) from 1 to 8.
We used seven models for these experiments: 
GPT-4o-2024-11-20,
Llama-3.1-70B-Instruct, 
Mistral-large, 
Llama-3.1-8B-Instruct, 
Granite-3.1-8B-Instruct, 
ToolACE-8B, 
and BitAgent-8B. 

\subsection{Challenge 3: Long Multi-Turn Conversations}\label{sec:ch_3_method}
% To test the effect of conversation length on information recall, we craft information, question pairs and inject multi-turn conversations of varying lengths between the information turn and the question. The number of tokens in the multi-turn conversation are controlled such that the conversation with token limit \( n_1 \) is a prefix of the conversation with token limit \( n_2 \) when \( n_1 \ < n_2 \).
% The multi-turn conversations are obtained from ComplexFuncBench~\citep{zhong2025complexfuncbench} and the information, question pairs are manually created. 

To construct a dataset for this evaluation, we leveraged multi-turn dialogs from ComplexFuncBench \citep{zhong2025complexfuncbench} and synthetically created two distinct settings: (1) the user introduces additional information (e.g., office location) at the start of the conversation within the user query and later requests a tool call that relies on that information (e.g., retrieving weather details for the office location), as illustrated in Figure \ref{fig:overview}; (2) similar to the first setting, but the required information is obtained through a tool call earlier in the conversation. The main distinction between (1) and (2) lies in how the necessary information is presented—setting (1) includes it in an unstructured format appended to the user query, while in setting (2), it is structured as a JSON-formatted response from a tool call. We applied both interaction types to 100 randomly selected conversations from ComplexFuncBench, resulting in 200 samples per token limit (10K, 20K, 40K, and 80K tokens).

% To construct a dataset for this evaluation, we leveraged multi-turn dialogs from ComplexFuncBench~\citep{zhong2025complexfuncbench} and artificially introduced an interaction type. Specifically, we modified the dialogs by having the user provide additional information (e.g., office location) at the beginning of the conversation and later request a tool call that utilizes this information (e.g., retrieving weather details for the specified location), as illustrated in Figure \ref{fig:overview}. This interaction type was applied to 100 sampled conversations from ComplexFuncBench, resulting in a dataset of 100 samples per token limit (10K, 20K, 40K, and 80K tokens), totaling 400 samples.

% Experimental setup
We assessed six models for this task: 
Llama-3.1-70B-Instruct, 
% Mistral-large,
DeepSeek-R1-Distill-Qwen-32B,
Llama-3.1-8B-Instruct, 
Granite-3.1-8B-Instruct, 
ToolACE-8B, 
and BitAgent-8B. 
To ensure consistency, we used BFCL's code for both the prompts and the output parser. Our evaluation focused on AST accuracy based on the BFCL metric.

%% file: sections/results.tex
\subsection{Challenge 1: Large Tool Catalog}
\begin{figure}[tb]
    \begin{subfigure}{0.32\textwidth}
    \includegraphics[width=\textwidth]{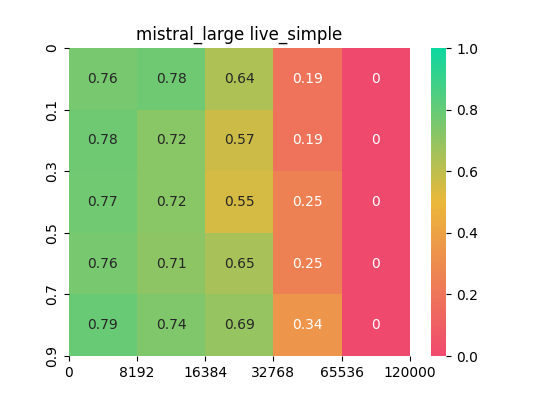}
       \caption{Mistral-large}
    \end{subfigure}
    \begin{subfigure}{0.32\textwidth}
    \includegraphics[width=\textwidth]{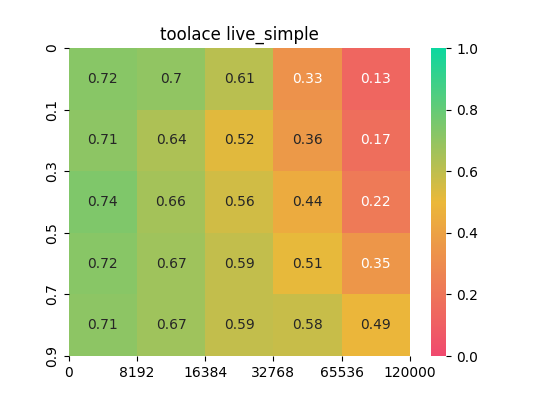}
       \caption{ToolACE-8B}
    \end{subfigure}
    \begin{subfigure}{0.32\textwidth}
    \includegraphics[width=\textwidth]{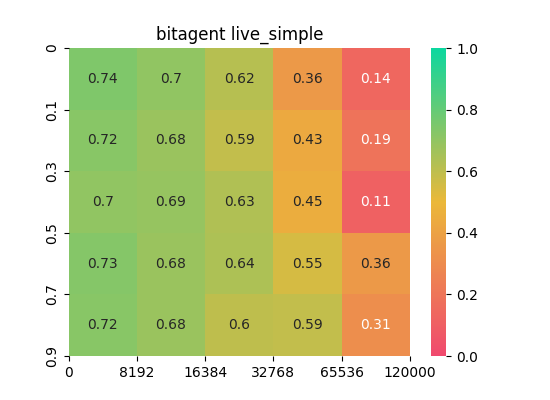}
       \caption{BitAgent-8B}
    \end{subfigure}

       \bigskip % <-- new 
       
    \begin{subfigure}{0.32\textwidth}
    \includegraphics[width=\textwidth]{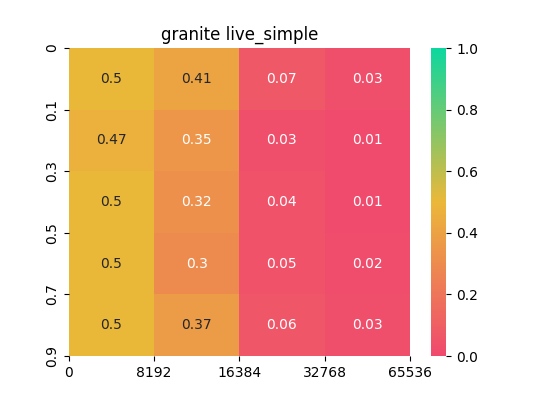}
       \caption{Granite-3.1-8b-instruct}
    \end{subfigure}
    \begin{subfigure}{0.32\textwidth}
    \includegraphics[width=\textwidth]{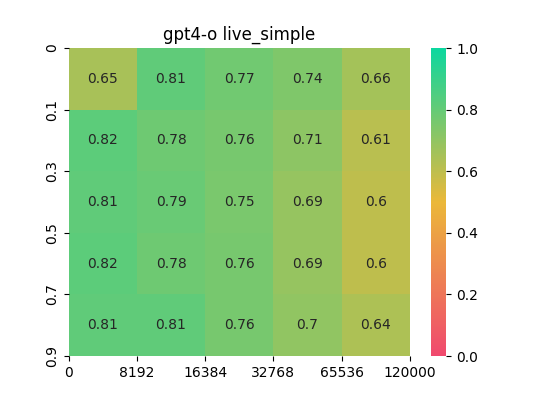}
       \caption{GPT-4o-2024-11-20}
    \end{subfigure}\hfill
    \begin{subfigure}{0.32\textwidth}
    \includegraphics[width=\textwidth]{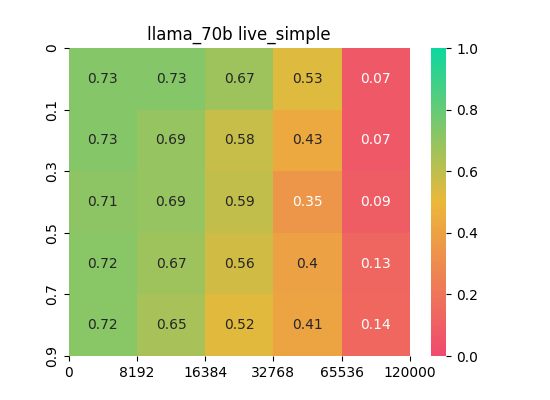}
       \caption{Llama-3.1-70B-instruct}
    \end{subfigure}

        \begin{subfigure}{0.32\textwidth}
    \includegraphics[width=\textwidth]{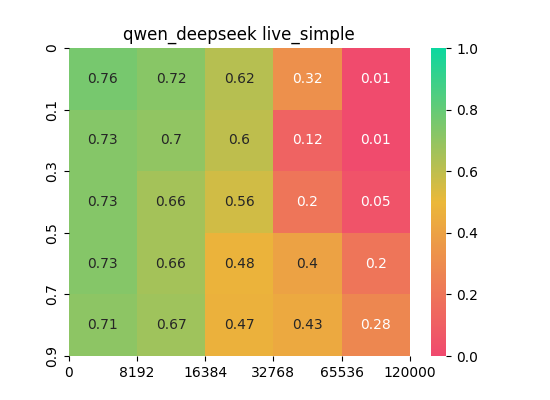}
       \caption{DeepSeek-R1-Distill-Qwen-32B}
    \end{subfigure}
    \begin{subfigure}{0.32\textwidth}
    \includegraphics[width=\textwidth]{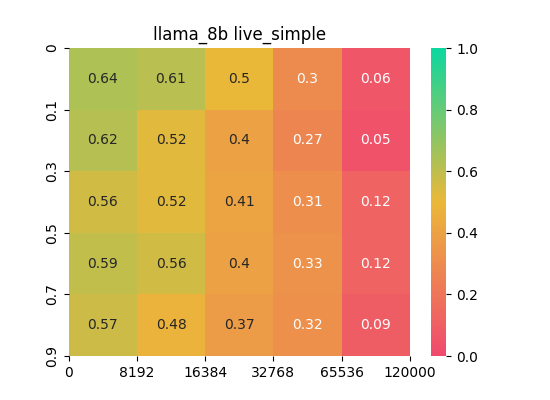}
       \caption{Llama-3.1-8B-Instruct}
    \end{subfigure}\hfill
    \begin{subfigure}{0.32\textwidth}
    \includegraphics[width=\textwidth]{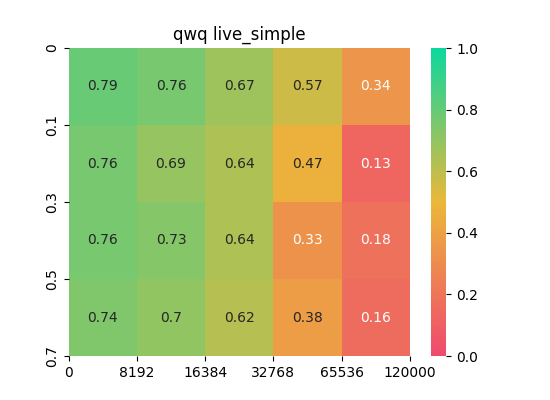}
       \caption{QwQ-32B}
    \end{subfigure}
    \caption{Challenge 1: AST accuracy for \texttt{live\_simple}. The performance degradation for larger contexts can be clearly seen as we move from left to right in each plot. For larger contexts, we can also notice a recency bias with the higher values of positions performing better.}
    \label{fig:c1_live_simple}
    \end{figure}

We plotted the AST accuracy of each model for all the three datasets as heatmaps. Figure \ref{fig:c1_live_simple} shows the results of for the \texttt{live\_simple} data subset of BFCL. The results for \texttt{simple} (figure \ref{fig:c1_simple}) and \texttt{multiple}  (figure \ref{fig:c1_multiple})  are in the appendix.  Across all the models (except for GPT-4o for \texttt{live\_simple} dataset), when the tool catalog is increased to reach a context length of 65,536 and 120,000, the performance deteriorates. The sensitivity to the position of the correct tool within the entire catalog also becomes more pronounced as the context length increases.

At context length 120,000, we observe that Mistral-large becomes very verbose and hallucinates the function names. Appendix \ref{appendix:error_analysis} includes some examples of this behavior. 
For context lengths 65536 and 120,000, ToolACE-8B, BitAgent-8B and Llama-3.1-70B-instruct tend to perform better when the tool occurs at later positions when compared to earlier. 
Granite-3.1-8B-instruct displays the characteristics of lost in the middle with the performance deteriorating when the tool occurs in the middle of the context. It performs well when it appears in the beginning or the ending of the context. GPT-4o generalizes well even for longer context lengths exhibiting a minimal drop in performance. 

% Among the three datasets, \emph{live simple} seems to be the hardest and \emph{simple} the easiest. 
Table \ref{tab:perf_drop} includes the minimum, maximum, and percentage degradation in AST accuracy for all the data subsets across context lengths and positions. Mistral-large has the largest percentage drop for all the datasets. 
For \texttt{simple} and \texttt{multiple}, BitAgent-8B has the second highest percentage drop and for \texttt{live\_simple}, Llama-3.1-70B-instruct did. We also analyze the performance drop across various context lengths for the same position and report that in Table \ref{tab:perf_drop_same_pos} in the Appendix. Across all the datasets, Mistral-large has the largest drop of 94\% and BitAgent-8B has the second highest with a drop of 82.37\%. We did a similar analysis along various positions for the same context length to analyze the position bias and observed that across all the datasets, BitAgent-8B has the highest and second highest drop of 83.47\% and 78.85\% respectively. This is reported in Table \ref{tab:perf_drop_same_con_len} of the Appendix. 

\begin{table}[t]
% \resizebox{1\textwidth}{!}{
\centering
\scriptsize
    \begin{tabular}{l|ccc|ccc|ccr}
    \toprule
    \multirow{2}{*}{\textbf{Model}} & \multicolumn{3}{|c|}{\textbf{Simple}} & \multicolumn{3}{|c|}{\textbf{Live Simple}} & \multicolumn{3}{|c}{\textbf{Multiple}}\\
    %\cline{2-10}
     & \textbf{Max} & \textbf{Min} & \textbf{\% Deg.} & \textbf{Max}  & \textbf{Min} & \textbf{\% Deg.} & \textbf{Max}  & \textbf{Min} & \textbf{\% Deg.} \\
     \midrule

Mistral-large & \textbf{0.94} & 0.00 & 94.00  & 0.78 & 0.00 & 78.29  & \textbf{0.94} & 0.00 & 94.00  \\ 
 % \hline 
 Granite-3.1-8B-instruct & 0.81 & 0.09 & 69.94  & 0.50 & 0.02 & 33.00  & 0.84 & 0.07 & 74.44  \\ 
 % \hline 
 ToolACE-8B & 0.92 & 0.28 & 62.17  & 0.72 & 0.13 & 34.19  & 0.94 & 0.25 & 63.82  \\ 
 % \hline 
 Llama 70B & 0.93 & 0.22 & 69.61  & 0.73 & 0.07 & 43.57  & 0.95 & 0.20 & 72.11  \\ 
 % \hline 
 BitAgent-8B & 0.92 & 0.12 & 78.96  & 0.74 & 0.14 & 57.82  & 0.95 & 0.10 & 82.37  \\ 
 % \hline 
 QwQ-32B & 0.91 & 0.45 & 40.89  & 0.79 & \textbf{0.34} & 21.17  & 0.92 & 0.47 & 42.05  \\ 
 % \hline 
GPT-4o-2024-11-20 & \textbf{0.94} & \textbf{0.76} & 12.88  & \textbf{0.82} & 0.60 & 10.62  & 0.94 & \textbf{0.78} & 13.81  \\ 
 % \hline 
 Llama-3.1-8B-Instruct & 0.86 & 0.09 & 76.11  & 0.64 & 0.06 & 50.74  & 0.86 & 0.07 & 75.83  \\ 
 % \hline 
  DeepSeek-R1-Distill-Qwen-32B & 0.92 & 0.02 & 89.30  & 0.76 & 0.01 & 73.41  & 0.89 & 0.01 & 85.96  \\ 
 % \hline 
 \bottomrule
    \end{tabular}
    \caption{Challenge 1 Performance Comparison: We compare the performance of all the models by varying the number of tools in the tool catalog across three different datasets (Simple, Live Simple and Multiple). The best accuracy (Max) and the worst accuracy (Min) for each model is reported, along with percentage degradation in performance. Most models tested show significant degradation in performance when the number of tools are increased.}
    \label{tab:perf_drop}
    % }
    \end{table}

\subsection{Challenge 2: Long Tool Responses}

\begin{figure}
    \centering
     \begin{subfigure}[b]{0.49\textwidth}
         \includegraphics[width=\textwidth]{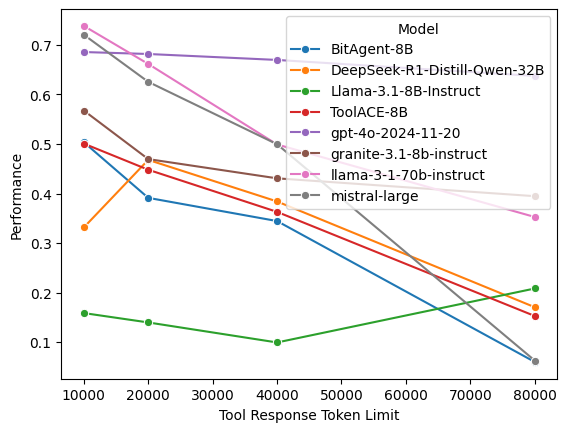}
         \caption{Tool Response Token Limit}
         \label{fig:rq2_plot}
     \end{subfigure}    
     \begin{subfigure}[b]{0.49\textwidth}
         \includegraphics[width=\textwidth]{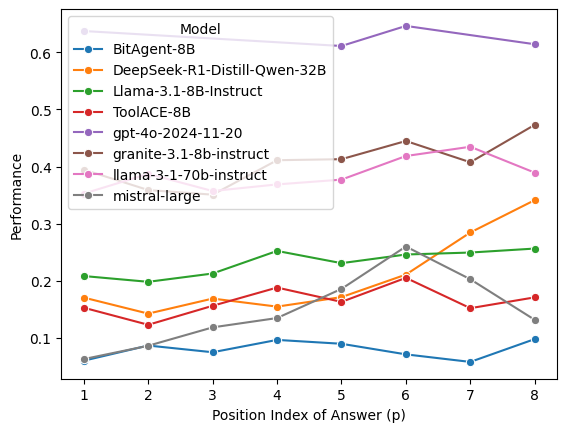}
         \caption{Answer Position}
         \label{fig:rq3_plot}
     \end{subfigure}    

    \caption{Challenge 2: Performance comparison (a) when tool response token limit is increased from 10K to 80K for position 1, (b) when position is varied from 1 to 8 for token limit of 80K. The performance degradation due to large responses is clearly seen in (a), whereas (b) shows recency bias.}
    \label{fig:task2_plots}
\end{figure}

For the long tool response challenge, Figure \ref{fig:rq2_plot} shows that the performance drops for most models including GPT-4o as the tool response size goes from 10K to 80K tokens. A degradation of 7\% for GPT-4o and a range of 91\% (Mistral-large) to 30.5\% (Granite-3.1-8b-instruct) for open-weight high performing models clearly confirms that LLMs struggle with long tool response understanding. Llama 8B is an exception to this trend and DeepSeek-R1-Distill-Qwen-32B performance for 10K is also an exception. For Llama 8B, we noticed that the model at shorter context lengths tries to generate json objects even though the task does not contain instructions to do so and this behavior is not observed at 80K context. In general, it can also be seen that the tool response QA as a task is not a trivial for LLMs, with the 10K token size performance ranging from 16\% to 74\% depending on the model. This challenges the current practice of appending the tool response to the current context for multi-turn tool calling, specially for large responses with complex structures. 
Appendix \ref{appendix:rq2_detailed_results} includes further analysis of model performance for question categories as described in section \ref{2:methodology}. Results show that some question categories (e.g., aggregation) are more challenging than others (e.g., extraction) where even GPT-4o experiences 10-20\% degradation on non-extractive tasks. 
% \textcolor{blue}{Should we write a sentence or two about conclusions from that?}

Figure \ref{fig:rq3_plot} shows how model performance varies with varying \( p \). Here, the number of tokens in the tool response is 80K for all values of \( p \). We observe that for all the models, the performance at \( p \) = 8 is higher than \( p \) = 1. This can be explained by recency bias~\citep{peysakhovich2023attentionsortingcombatsrecency} where models are better at attending recent tokens. Overall, the performance variation based on position ranges from 5\% (GPT-4o) to 75\% (Mistral-large) depending on the model. 

\subsection{Challenge 3: Long Multi-Turn Conversations}

\begin{table}[t]
\centering
\resizebox{\textwidth}{!}{%
\begin{tabular}{l|cccc|cccc}
\toprule
\multicolumn{1}{c|}{\multirow{2}{*}{\textbf{Models}}} & \multicolumn{4}{c|}{\textbf{Setting - 1 (unstructured)}}                          & \multicolumn{4}{c}{\textbf{Setting - 2 (structured)}}                             \\ 
\multicolumn{1}{c|}{}                                 & \textbf{10K } & \textbf{20K } & \textbf{40K } & \textbf{80k } & \textbf{10K } & \textbf{20K } & \textbf{40K } & \textbf{80k } \\ \midrule
{ToolACE-8B}                                   & 0.24               & 0.15               & 0.11               & 0.10               & \textbf{0.91 }              & \textbf{0.74}               & \textbf{0.67}               & 0.29               \\
{BitAgent-8B}                                  & 0.49               & 0.41               & 0.24               & 0.28               & 0.87               & 0.66               & 0.51               & \textbf{0.32}               \\
{Granite-3.1-8B-Instruct}                      & 0.15               & 0.06               & 0.07               & 0.05               & 0.82               & 0.54               & 0.24               & 0.04               \\
{Llama-3.1-8B-Instruct}                        & 0.10               & 0.22               & 0.24               & 0.24               & 0.22               & 0.18               & 0.23               & 0.10               \\
{DeepSeek-R1-Distill-Qwen-32B}                 & 0.40               & 0.07               & 0.12               & 0.10               & 0.23               & 0.03               & 0.07               & 0.02               \\
{Llama-3.1-70B-Instruct}                       & \textbf{0.96 }              & \textbf{0.98 }              & \textbf{0.99 }              & \textbf{0.99 }              & 0.38               & 0.31               & 0.35               & 0.19    \\
\bottomrule
\end{tabular}%
}
\caption{Challenge 3 Evaluation Results; all the scores are \textit{AST} accuracies. The accuracies drop significantly as we move from 10K to 80K token conversations for most models.}
\label{tab:task_3_results}
\end{table}

Finally, for the multi-turn challenge, Table \ref{tab:task_3_results} summarizes the evaluation results for all models described in Section \ref{sec:ch_3_method}. It reports AST match accuracy across context lengths of 10K, 20K, 40K, and 80K tokens for both structured and unstructured settings. The results show that most models experience a drop in accuracy as the context size increases in both settings. However, the Llama models show stable performance on setting-1 (unstructured), with Llama-3.1-70B-Instruct outperforming all others—achieving over 95\% accuracy across all context lengths. In contrast, ToolACE-8B and BitAgent-8B perform well on setting-2 (structured), though their accuracy declines with larger contexts. We observed that smaller models like ToolACE-8B, BitAgent-8B, and Granite-3.1-8B-Instruct perform better in the structured setting than in the unstructured one. Conversely, larger models such as DeepSeek-R1-Distill-Qwen-32B and Llama-3.1-70B-Instruct demonstrate stronger performance in the unstructured setting compared to the structured.

%% file: sections/conclusion.tex
In this paper, we studied the three main challenges that LLMs face in the long context tool calling scenario. For the large tool catalog challenge, we evaluate LLMs by gradually increasing the number of tools and varying the position of the correct tool within the catalog. To examine LLMs' understanding of long tool responses, we construct a dataset by leveraging public APIs and vary the lengths of the responses. For the long multi-turn conversations, we evaluate the long dependency challenge due to a large number of conversation turns. Across all these tasks, we notice that LLMs' performance deteriorated significantly as the context length increases. For challenge 1, we also observe that the position bias is more pronounced as the context length increases. 

Future work will include diving deeper into the reasons behind this decline in performance by further analyzing the errors we encountered. We also plan on expanding the datasets to include more variations in each task. Finally, we plan on extending the analysis of Challenge 3 to structured text as well.

%% file: sections/appendix.tex
% \documentclass{article}
% \usepackage[submission]{colm2025_conference}
% \usepackage{todonotes}

% \usepackage{graphicx} % Required for inserting images
% \usepackage{lineno}

% \title{LCTC: A Study of Long Context Abilities of LLMs for Tool Calling}
% %\author{kakate }
\appendix
\label{appendix}
\section{Data Details}
\label{appendix:data_details}
\subsection{Challenge 1}
\subsubsection{Tool Catalog details}
Table~\ref{tab:num_tools} shows the average number of tools across all the positions and all the datasets for each context length.

\begin{table}[h]
    \centering
    \begin{tabular}{rr}
    \toprule
        Context Length & \# Tools \\
        \midrule
        8,192 & 49 \\
        16,384 & 102 \\
        32,768 & 207 \\
        65,536 & 417 \\
        120,000 & 741 \\
\bottomrule                
    \end{tabular}
    \caption{Number of tools per context length}
    \label{tab:num_tools}
\end{table}

\subsection{Challenge 2}

% \begin{figure}[!h]
%     \centering
%     \includegraphics[width=1\linewidth]{sections/figures/Task2_motivation.png}
%     \caption{Motivation for Challenge 2}
%     \label{fig:task2_motive}
% \end{figure}

Table \ref{tab:tool_response_qa_data_stats} shows the number of QA samples per endpoint for the dataset used for Challenge 2. The API endpoints are from the ComplexFuncBench dataset released by ~\citep{zhong2025complexfuncbench}.

\begin{table}[h!]
    \centering
    \begin{tabular}{lrrrr}
    \toprule
        Endpoint & \# Total  & \# Extraction & \# Filtering & \# Aggregation\\
        \midrule
        Search\_Flights\_Multi\_Stops & 152 & 100 & 26 & 26\\
        Get\_Availability & 132 & 47 & 39 & 46\\
        Get\_Room\_List\_With\_Availability & 126 & 96 & 13 & 17\\
        Get\_Seat\_Map & 96 & 16 & 40 & 40\\
        Search\_Car\_Rentals & 60 & 20 & 20 & 20\\
\bottomrule        
    \end{tabular}
    \caption{Data distribution for Challenge 2}
    \label{tab:tool_response_qa_data_stats}
\end{table}

% \begin{tabular}{l|ccc|ccc}
% \toprule

% \multirow{2}{*}{\textbf{Model}} & \multicolumn{3}{c|}{\textbf{Context Length}} & \multicolumn{3}{c}{\textbf{Answer Position}} \\

%  & \textbf{Min} & \textbf{Max} & \textbf{\% Degradation} & \textbf{Min} & \textbf{Max} & \textbf{\% Degradation} \\
% \midrule

\section{Result Details}
\subsection{Challenge 1}

Table \ref{tab:perf_drop_same_pos} shows the minimum, maximum, and percentage degradation in AST accuracy for position 0.1 and different context lengths for different data splits from BFCL.

Table \ref{tab:perf_drop_same_con_len} shows the minimum, maximum, and percentage degradation in AST accuracy for context length 120K, but varying the position values. 

\begin{figure}[h!]
    \begin{subfigure}{0.32\textwidth}
    \includegraphics[width=\textwidth]{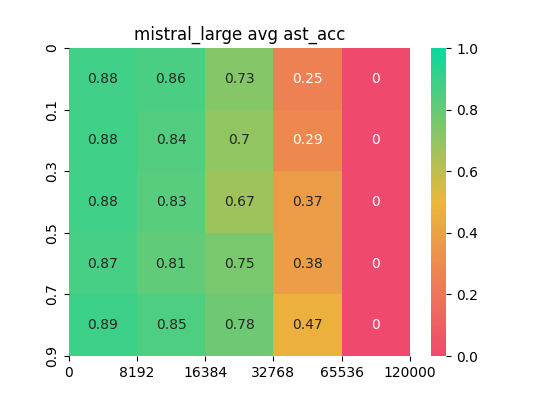}
       \caption{Mistral-Large-Instruct-2411}
    \end{subfigure}
    \begin{subfigure}{0.32\textwidth}
    \includegraphics[width=\textwidth]{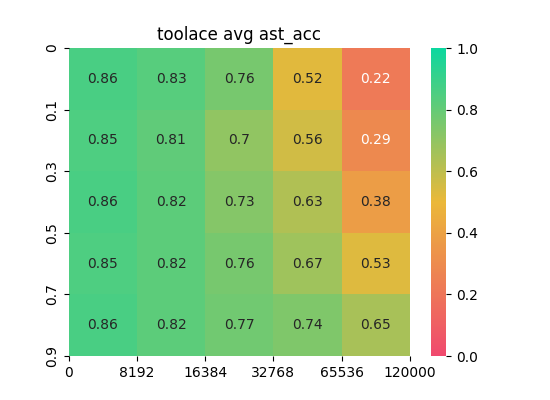}
       \caption{ToolACE-8B}
    \end{subfigure}
    \begin{subfigure}{0.32\textwidth}
    \includegraphics[width=\textwidth]{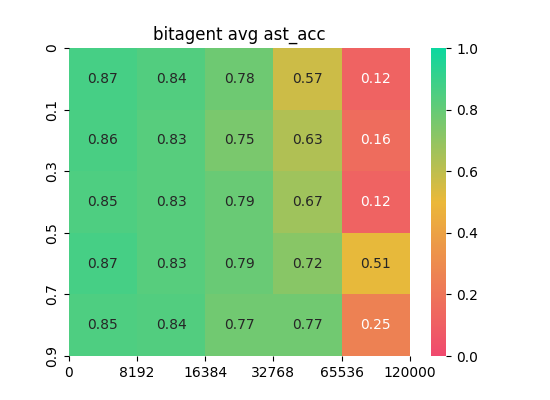}
       \caption{BitAgent-8B}
    \end{subfigure}

       \bigskip % <-- new 
       
    \begin{subfigure}{0.32\textwidth}
    \includegraphics[width=\textwidth]{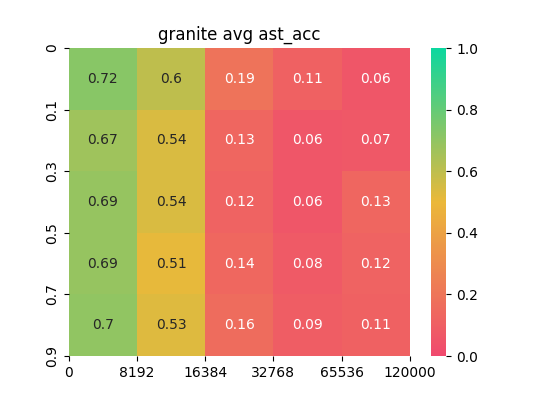}
       \caption{granite-3.1-8b-Instruct}
    \end{subfigure}
    \begin{subfigure}{0.32\textwidth}
    \includegraphics[width=\textwidth]{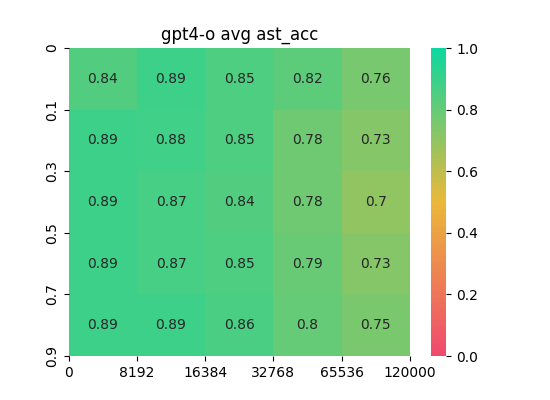}
       \caption{gpt-4o-2024-11-20}
    \end{subfigure}\hfill
    \begin{subfigure}{0.32\textwidth}
    \includegraphics[width=\textwidth]{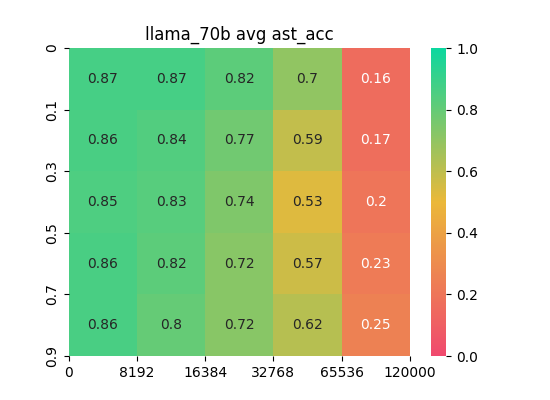}
       \caption{llama-3-1-70B-Instruct}
    \end{subfigure}

   \begin{subfigure}{0.32\textwidth}
    \includegraphics[width=\textwidth]{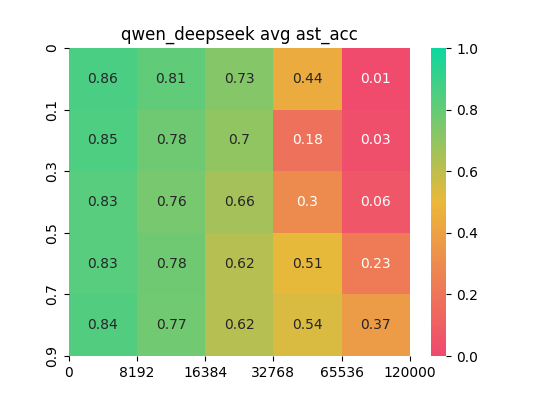}
       \caption{DeepSeek-R1-Distill-Qwen-32B}
    \end{subfigure}
    \begin{subfigure}{0.32\textwidth}
    \includegraphics[width=\textwidth]{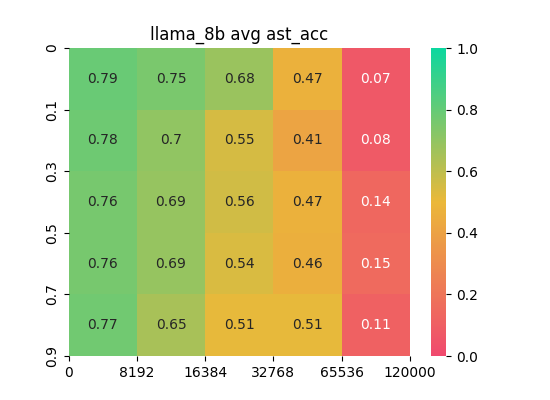}
       \caption{Llama-3.1-8B-Instruct}
    \end{subfigure}\hfill
    \begin{subfigure}{0.32\textwidth}
    \includegraphics[width=\textwidth]{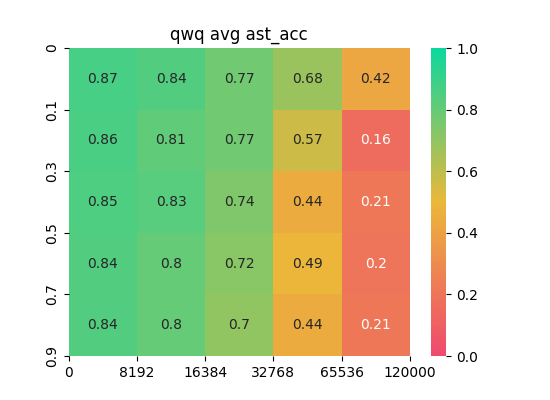}
       \caption{QwQ-32B}
    \end{subfigure}
    \caption{Challenge 1: average AST accuracy for all datasets}
    \label{fig:c1_avg}
    \end{figure}

Figure \ref{fig:c1_avg} shows the variation in the average AST accuracy across the subsets simple, live\_simple, and multiple for different models.

\begin{figure}[h!]
    \begin{subfigure}{0.32\textwidth}
    \includegraphics[width=\textwidth]{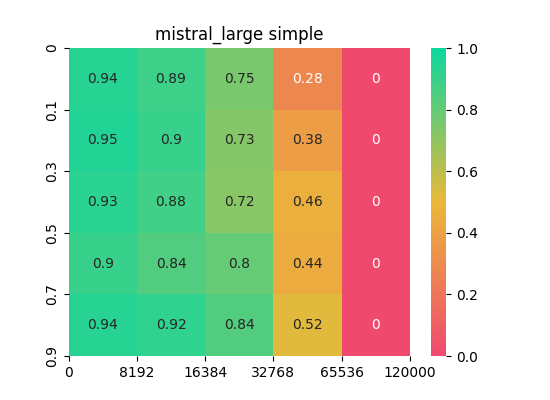}
       \caption{Mistral-Large-Instruct-2411}
    \end{subfigure}
    \begin{subfigure}{0.32\textwidth}
    \includegraphics[width=\textwidth]{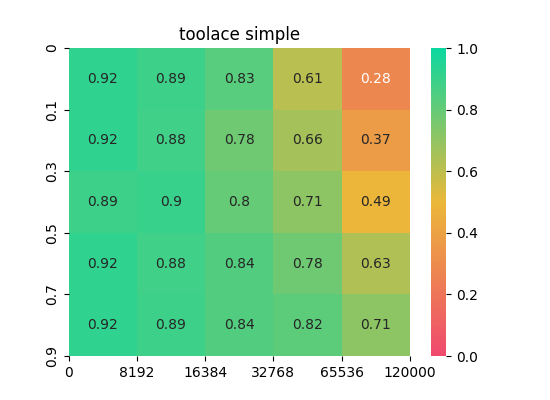}
       \caption{ToolACE-8B}
    \end{subfigure}
    \begin{subfigure}{0.32\textwidth}
    \includegraphics[width=\textwidth]{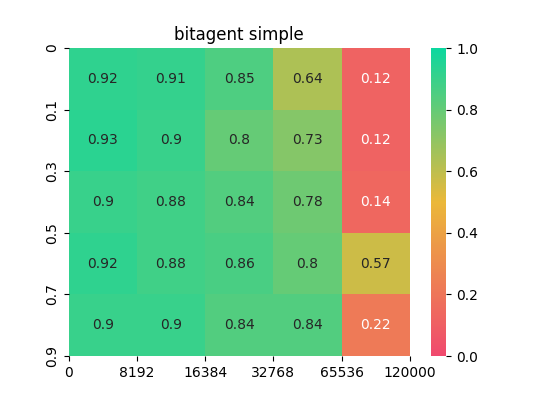}
       \caption{BitAgent-8B}
    \end{subfigure}

       \bigskip % <-- new 
       
    \begin{subfigure}{0.32\textwidth}
    \includegraphics[width=\textwidth]{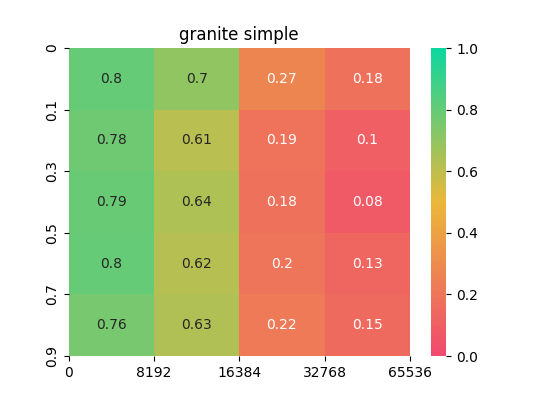}
       \caption{granite-3.1-8B-Instruct}
    \end{subfigure}
    \begin{subfigure}{0.32\textwidth}
   \includegraphics[width=\textwidth]{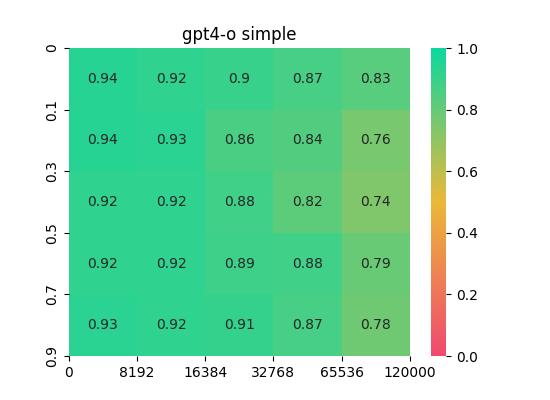}
       \caption{gpt-4o-2024-11-20}
    \end{subfigure}\hfill
    \begin{subfigure}{0.32\textwidth}
    \includegraphics[width=\textwidth]{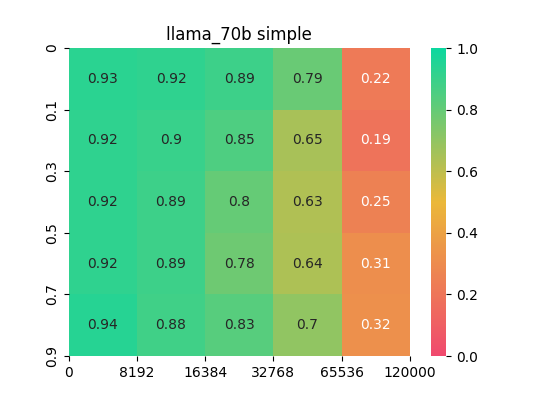}
       \caption{llama-3-1-70B-Instruct}
    \end{subfigure}

 %    \bigskip % <-- new
        \begin{subfigure}{0.32\textwidth}
    \includegraphics[width=\textwidth]{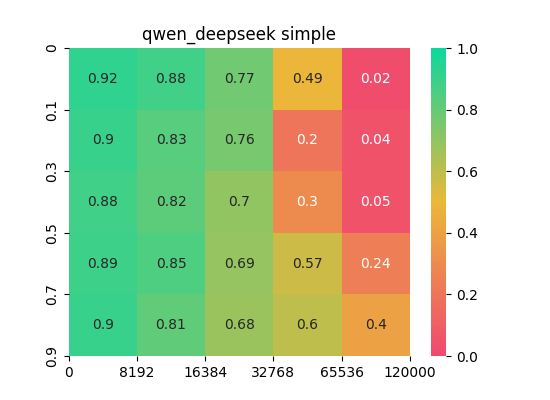}
       \caption{DeepSeek-R1-Distill-Qwen-32B}
    \end{subfigure}
    \begin{subfigure}{0.32\textwidth}
    \includegraphics[width=\textwidth]{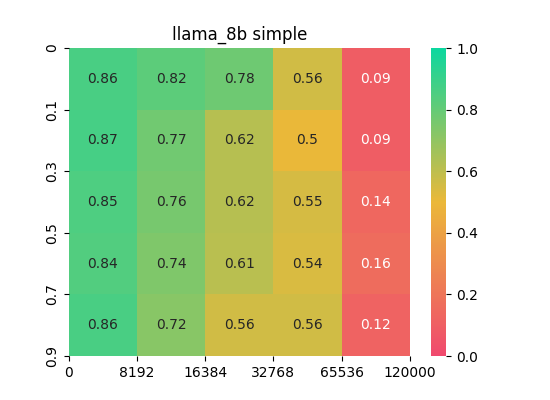}
       \caption{Llama-3.1-8B-Instruct}
    \end{subfigure}\hfill
    \begin{subfigure}{0.32\textwidth}
    \includegraphics[width=\textwidth]{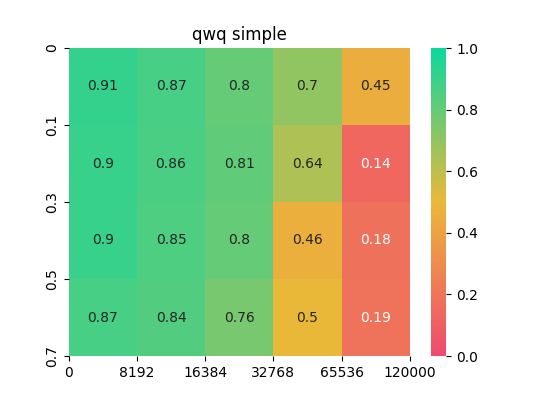}
       \caption{QwQ-32B}
    \end{subfigure}
    \caption{Challenge 1: AST accuracy for the simple subset}
    \label{fig:c1_simple}
    \end{figure}

\begin{figure}[h!]
    \begin{subfigure}{0.32\textwidth}
    \includegraphics[width=\textwidth]{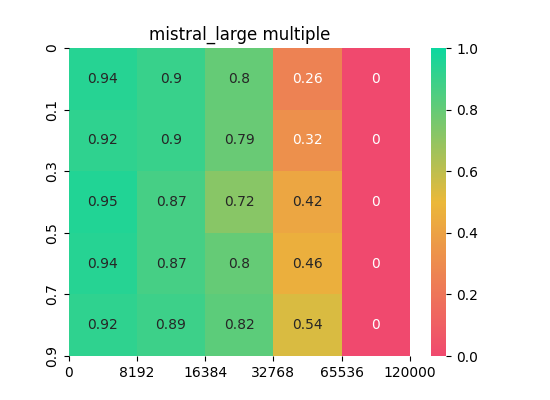}
       \caption{Mistral-Large-Instruct-2411}
    \end{subfigure}
    \begin{subfigure}{0.32\textwidth}
    \includegraphics[width=\textwidth]{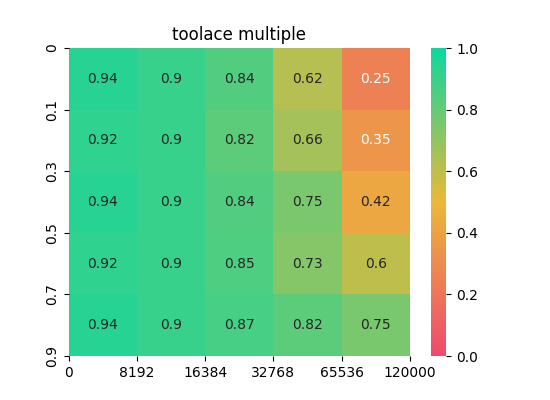}
       \caption{ToolACE-8B}
    \end{subfigure}
    \begin{subfigure}{0.32\textwidth}
    \includegraphics[width=\textwidth]{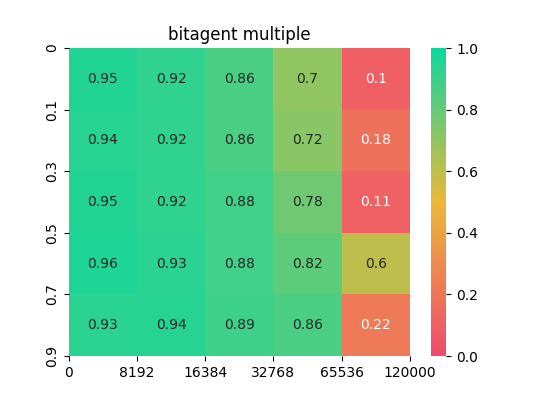}
       \caption{BitAgent-8B}
    \end{subfigure}

       \bigskip % <-- new 
       
    \begin{subfigure}{0.32\textwidth}
    \includegraphics[width=\textwidth]{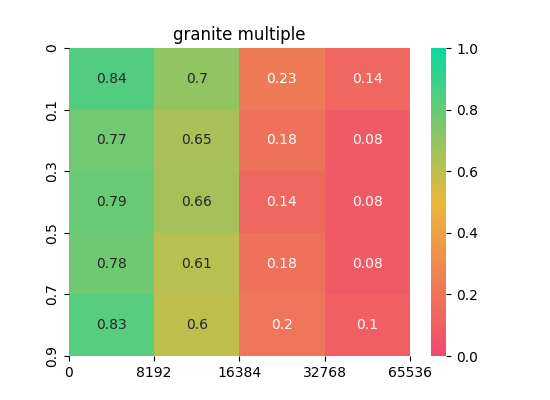}
       \caption{granite-3.1-8B-Instruct}
    \end{subfigure}
    \begin{subfigure}{0.32\textwidth}
   \includegraphics[width=\textwidth]{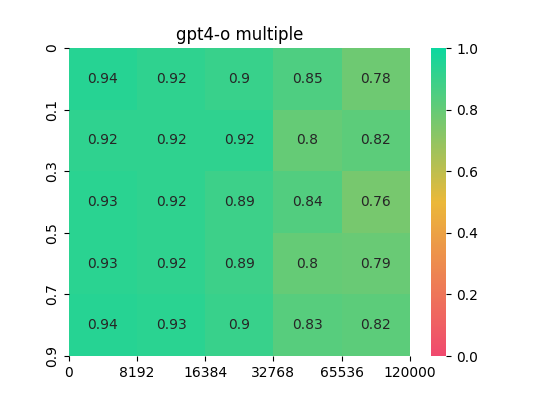}
       \caption{gpt-4o-2024-11-20}
    \end{subfigure}\hfill
    \begin{subfigure}{0.32\textwidth}
    \includegraphics[width=\textwidth]{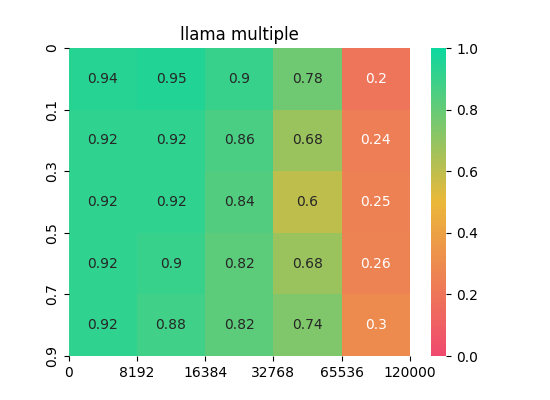}
       \caption{llama-3-1-70B-Instruct}
    \end{subfigure}

    \bigskip % <-- new
       \begin{subfigure}{0.32\textwidth}
    \includegraphics[width=\textwidth]{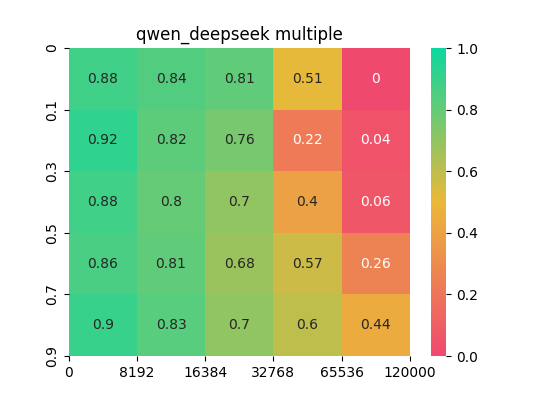}
       \caption{DeepSeek-R1-Distill-Qwen-32B}
    \end{subfigure}
    \begin{subfigure}{0.32\textwidth}
    \includegraphics[width=\textwidth]{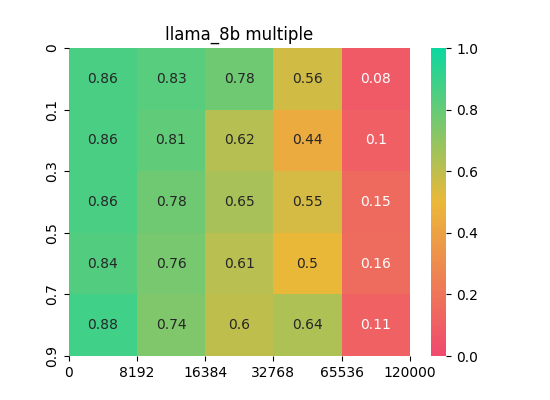}
       \caption{Llama-3.1-8B-Instruct}
    \end{subfigure}\hfill
    \begin{subfigure}{0.32\textwidth}
    \includegraphics[width=\textwidth]{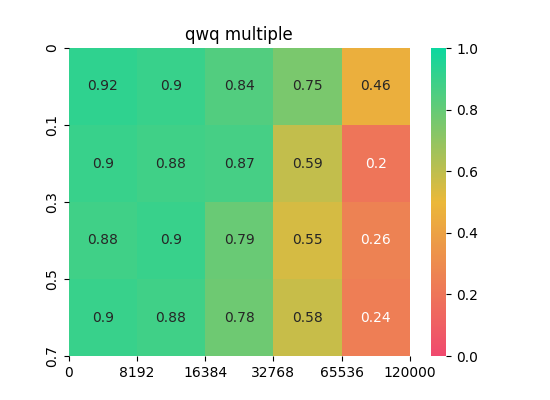}
       \caption{QwQ-32B}
    \end{subfigure}
    \caption{Challenge 1: average AST accuracy the multiple subset}
    \label{fig:c1_multiple}
    \end{figure}

Figures \ref{fig:c1_simple} and \ref{fig:c1_multiple} show the variation in the AST accuracy for different models for the subsets simple and multiple respectively.

\begin{table}[h!]
\centering
    \scriptsize
    \begin{tabular}{l|ccr|ccr|ccr}
    \toprule
    \multirow{2}{*}{\textbf{Model}} & \multicolumn{3}{|c|}{\textbf{Simple}} & \multicolumn{3}{|c|}{\textbf{Live Simple}} & \multicolumn{3}{|c}{\textbf{Multiple}}\\
    
     & \textbf{Max} & \textbf{Min} & \textbf{\% Deg.} & \textbf{Max}  & \textbf{Min} & \textbf{\% Deg.} & \textbf{Max}  & \textbf{Min} & \textbf{\% Deg.} \\
    \midrule
Mistral Large & 0.94 & 0.00 & 94.00  & 0.78 & 0.00 & 78.29  & 0.94 & 0.00 & 94.00  \\ 
 
Granite-3.1-8B-inst & 0.81 & 0.09 & 69.94  & 0.50 & 0.02 & 33.00  & 0.84 & 0.07 & 74.44  \\ 
 
 ToolACE-8B & 0.92 & 0.28 & 62.17  & 0.72 & 0.13 & 34.19  & 0.94 & 0.25 & 63.82  \\ 
 
 Llama-3.1-70B-inst  & 0.93 & 0.22 & 69.61  & 0.73 & 0.07 & 43.57  & 0.95 & 0.20 & 72.11  \\ 
 
 BitAgent-8B & 0.92 & 0.12 & 78.96  & 0.74 & 0.14 & 57.82  & 0.95 & 0.10 & 82.37  \\ 
  
 QwQ-32B & 0.91 & 0.45 & 40.89  & 0.79 & 0.34 & 21.17  & 0.92 & 0.47 & 42.05  \\ 
  
 GPT-4o-2024-11-20  & 0.94 & 0.83 & 4.95  & 0.81 & 0.60 & -20.88  & 0.94 & 0.78 & 6.40  \\ 

 Llama-3.1-8B-inst & 0.86 & 0.09 & 76.11  & 0.64 & 0.06 & 50.74  & 0.86 & 0.07 & 75.83  \\ 

 DeepSeek-R1-Distill-Qwen-32B & 0.92 & 0.02 & 89.30  & 0.76 & 0.01 & 73.41  & 0.89 & 0.01 & 85.96  \\
 
 \bottomrule
    \end{tabular}
    \caption{Performance drop for same position}
    \label{tab:perf_drop_same_pos}

    \end{table}

\begin{table}[h!]
\centering
    \scriptsize
    \begin{tabular}{l|ccr|ccr|ccr}
    \toprule
    \multirow{2}{*}{\textbf{Model Name}} & \multicolumn{3}{|c|}{\textbf{Simple}} & \multicolumn{3}{|c|}{\textbf{Live Simple}} & \multicolumn{3}{|c}{\textbf{Multiple}}\\

     & \textbf{Max} & \textbf{Min} & \textbf{\% Deg.} & \textbf{Max}  & \textbf{Min} & \textbf{\% Deg.} & \textbf{Max}  & \textbf{Min} & \textbf{\% Deg.} \\
    \midrule
 Mistral Large & 0.53 & 0.28 & 46.67  & 0.34 & 0.19 & 46.07  & 0.54 & 0.27 & 50.47  \\ 
 
 Granite-3.1-8B-inst & 0.18 & 0.08 & 55.56  & 0.06 & 0.01 & 80.00  & 0.17 & 0.07 & 62.86  \\ 
 
 ToolACE-8B & 0.71 & 0.28 & 60.92  & 0.49 & 0.13 & 73.23  & 0.75 & 0.25 & 66.67  \\ 
 
 Llama-3.1-70B-inst & 0.32 & 0.19 & 40.48  & 0.14 & 0.07 & 47.22  & 0.29 & 0.20 & 33.90  \\ 
 
 BitAgent-8B & 0.57 & 0.12 & 78.85  & 0.36 & 0.11 & 69.57  & 0.60 & 0.10 & 83.47  \\ 
 
 QwQ-32B & 0.45 & 0.14 & 67.96  & 0.34 & 0.13 & 61.36  & 0.47 & 0.20 & 58.06  \\ 
 
 GPT-4o-2024-11-20  & 0.83 & 0.74 & 11.41  & 0.81 & 0.65 & 20.00  & 0.82 & 0.76 & 7.88  \\ 

 Llama-3.1-8B-inst & 0.17 & 0.09 & 46.97  & 0.12 & 0.05 & 53.33  & 0.15 & 0.07 & 51.61  \\ 

 DeepSeek-R1-Distill-Qwen-32B & 0.4 & 0.02 & 94.30  & 0.28 & 0.01 & 97.22  & 0.45 & 0.01 & 98.88  \\ 
 \bottomrule
 \end{tabular}
 \caption{Performance drop for same context}
 \label{tab:perf_drop_same_con_len}
    \end{table}

\subsection{Challenge 2}

\label{appendix:rq2_detailed_results}
Table \ref{tab:rq2_rq3_results} includes the minimum, maximum, and percent degradation for each model for the two experiments in challenge 2. 

\begin{figure}[h!]
    \centering
     \begin{subfigure}[b]{0.47\textwidth}
         \includegraphics[width=\textwidth]{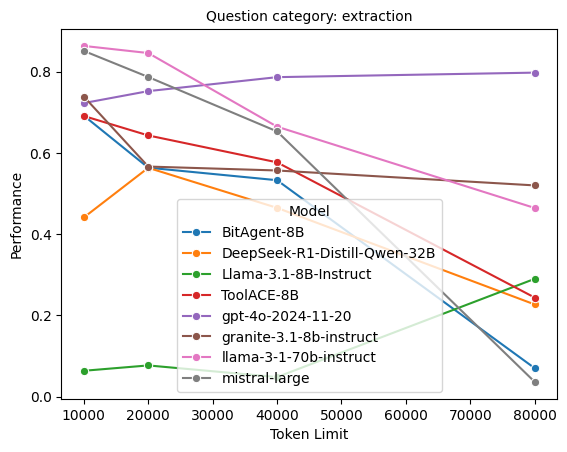}
         \caption{Extraction}
         \label{fig:rq2_extraction}
     \end{subfigure}    
     \begin{subfigure}[b]{0.47\textwidth}
         \includegraphics[width=\textwidth]{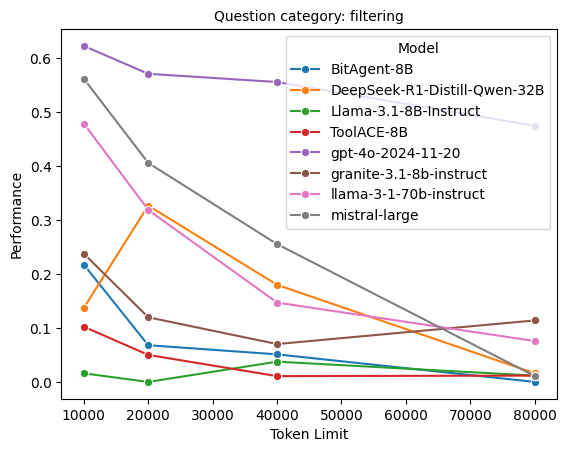}
         \caption{Filtering}
         \label{fig:rq2_filtering}
     \end{subfigure}    
     \begin{subfigure}[b]{0.47\textwidth}
         \includegraphics[width=\textwidth]{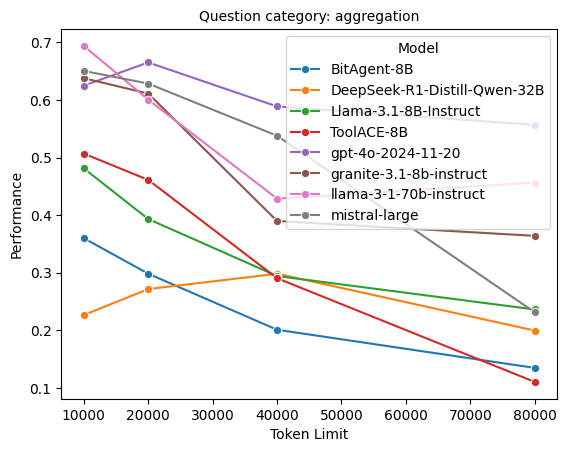}
         \caption{Aggregation}
         \label{fig:rq2_aggregation}
     \end{subfigure}    
     \begin{subfigure}[b]{0.47\textwidth}
         \includegraphics[width=\textwidth]{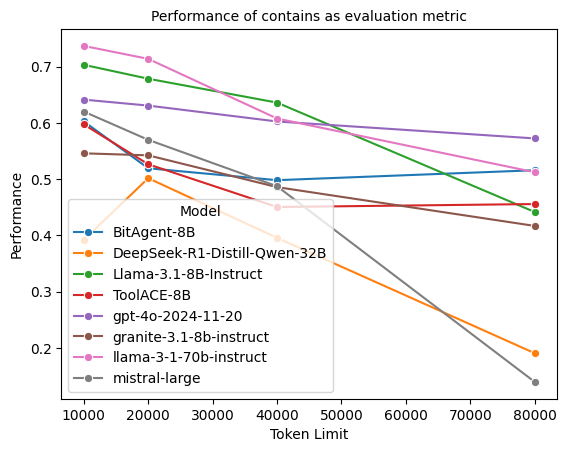}
         \caption{Contains}
         \label{fig:rq2_contains}
     \end{subfigure}    

    \caption{Drill-down of challenge 2 results based on question categories and performance based on the 'contains' metric}
    \label{fig:task2_plots_app}
\end{figure}

Figure \ref{fig:task2_plots_app} shows the plots for each question category and a plot of the "contains" metric.

\begin{table}[h]
\centering
\begin{tabular}{l|ccc|ccc}
% \small
\toprule

\multirow{2}{*}{\textbf{Model}} & \multicolumn{3}{c|}{\textbf{Context Length}} & \multicolumn{3}{c}{\textbf{Answer Position}} \\

 & \textbf{Min} & \textbf{Max} & \textbf{\% Degradation} & \textbf{Min} & \textbf{Max} & \textbf{\% Degradation} \\
\midrule
BitAgent-8B              & 0.06 & 0.50 & 88.11 & 0.06 & 0.10 & 40.95 \\
ToolACE-8B               & 0.15 & 0.50 & 69.52 & 0.12 & 0.21 & 40.12 \\
GPT-4o-2024-11-20        & 0.64 & 0.69 &  7.04 & 0.61 & 0.65 &  5.44 \\
Granite-3.1-8B-instruct  & 0.39 & 0.57 & 30.47 & 0.35 & 0.47 & 25.87 \\
Llama-3.1-70B-instruct   & 0.35 & 0.74 & 52.25 & 0.35 & 0.43 & 18.89 \\
Mistral-large            & 0.06 & 0.72 & 91.30 & 0.06 & 0.26 & 75.88 \\
DeepSeek-R1-Distill-Qwen-32B & 0.17 & 0.47 & 63.8 & 0.14 & 0.34 & 58.82\\
Llama-3.1-8B-Instruct & 0.1 & 0.21 & 52.38 & 0.20 & 0.26 & 23.07 \\
\bottomrule
\end{tabular}
\caption{Challenge 2 raw results}
\label{tab:rq2_rq3_results}
\end{table}

\newpage
\section{Models and Prompts}

\subsection{Models}
\label{appendix:models}
Table \ref{tab:models} lists all the models used in this study, with links to where each model can be found online.

\begin{table}
    \centering
    \scriptsize
    \begin{tabular}{ll}
    \toprule
        Model & Model Pointer \\
        \midrule
        BitAgent-8B & https://huggingface.co/BitAgent/BitAgent-8B \\
        GPT-4o-2024-11-20 & https://openai.com/index/hello-gpt-4o/ \\
        Granite-3.1-8b-instruct & https://huggingface.co/ibm-granite/granite-3.1-8b-instruct \\
        Llama-3-1-70b-instruct & https://huggingface.co/meta-llama/Llama-3.1-70B-Instruct \\
        Llama-3.1-8B-Instruct  & https://huggingface.co/meta-llama/Llama-3.1-8B-Instruct \\
        Mistral-large & https://huggingface.co/mistralai/Mistral-Large-Instruct-2411  \\
        ToolACE-8B & https://huggingface.co/Team-ACE/ToolACE-8B \\
        DeepSeek-R1-Distill-Qwen-32B & https://huggingface.co/deepseek-ai/DeepSeek-R1-Distill-Qwen-32B \\
        QwQ-32B & https://huggingface.co/Qwen/QwQ-32B \\
     \bottomrule           
    \end{tabular}
    \caption{A list of all the models used in our study}
    \label{tab:models}
\end{table}

\subsection{Prompts}

\textbf{Challenge 1}: For these experiments we used the prompts from the Berkeley Function Calling Leaderboard~\citep{berkeley-function-calling-leaderboard} as is. 

\textbf{Challenge 2}:

\begin{lstlisting}[basicstyle=\ttfamily, breaklines=true]
------------------------------------------------------------------
Zero-shot prompt for tool response QA: 
------------------------------------------------------------------

You are given a response from an API call (in JSON format).
Answer the question based on the information provided in the API response.

```json
{api_response}
```

Question: {question}

Only respond with the answer. Do not include any other text or json in the response.
Do not rephrase the answer or write it in complete sentence, return exactly as is from the JSON.

Answer:
------------------------------------------------------------------
\end{lstlisting}

\section{Error Analysis}
\label{appendix:error_analysis}

\subsection{Challenge 1: Large Tool Catalog}

\textbf{Types of Errors:} As noted above, the number of errors increases with an increase in context size, for all models. 
For most models, the most common errors are (1) invalid formatting (AST errors) (2) calling the wrong function and (3) calling the right function with the wrong parameters. There is a cascading effect, where models which perform poorly often make more AST errors, while the models which perform better often have the right formatting, but make errors in the function names and errors. It is interesting to note that even models that perform reasonably well as the context size increases (like GPT-4o, for example), also make formatting errors, albeit they are not most common type of errors. Similarly, when the context length is small, models make fewer formatting errors, and the most common errors are more subtle (for example, passing the wrong parameter). 

\textbf{Examples of Failures:} The responses of Mistral large collapsed altogether for context length 120k. The listing below shows some sample responses for the \texttt{simple} dataset. For example, the response for query id simple\_22 is very verbose and does not return a function name. The response for query id simple\_17 returns prime\_factors but the correct response is get\_prime\_factors.

\begin{lstlisting}[basicstyle=\ttfamily, breaklines=true]
{"id": "simple_0", "result": " To calculate the area of a triangle, you can use the formula:\n\n\\[ \\text{Area} = \\frac{1}{2} \\times \\text{base} \\times \\text{height} \\]\n\nSubstituting the given values:\n\n\\[ \\text{base} = 10 \\]\n\\[ \\text{height} = 5 \\]\n\n\\[ \\text{Area} = \\frac{1}{2} \\times 10 \\times 5 = 25 \\]\n\nSo, the area of the triangle is 25 square units.", "input_token_count": 127315, "output_token_count": 126, "latency": 32.22818660736084}

{"id": "simple_1", "result": " To calculate the factorial of 5, we can use the formula:\n\n5! = 1 * 2 * 3 * 4 * 5\n\nHere's the breakdown of the calculation:\n\n1. Calculate the product of the first 5 positive integers:\n   1 * 2 * 3 * 4 * 5 = 120\n\n2. Calculate the product of the next 5 positive integers:\n   2 * 3 * 4 * 5 = 240\n\n3. Calculate the product of the next 5 positive integers:\n   3 * 4 * 5 = 360\n\n4. Calculate the product of the next 5 positive integers:\n   4 * 5 = 480\n\n5. Calculate the product of the next 5 positive integers:\n   5 * 6 * 7 * 8 * 9 * 10 = 5760\n\nSo, the factorial of 5 is 5760.", "input_token_count": 127365, "output_token_count": 235, "latency": 35.642239809036255}

{"id": "simple_17", "result": " {'name': 'prime_factors', 'description': 'Find the prime factors of a given number. Note that the provided function is in Python 3 syntax.', 'parameters': {'type': 'dict', 'properties': {'number': {'type': 'integer', 'description': 'The number to find the prime factors of.'}}, 'required': ['number']}}", "input_token_count": 127453, "output_token_count": 77, "latency": 32.36869549751282}

{"id": "simple_22", "result": " To do this, we need to find the greatest common divisor (GCD) of the two numbers. The GCD of two numbers a and b is the largest number that divides both a and b without leaving a remainder.\n\nHere's the step-by-step process to calculate the GCD of 12 and 15:\n\n1. Start with the larger number (15) and subtract the smaller number (12) repeatedly until the remainder is less than the smaller number.\n2. The first subtraction gives 15 - 12 = 3.\n3. The second subtraction gives 3 - 12 = 1.\n4. The third subtraction gives 1 - 12 = -1.\n5. The fourth subtraction gives -1 - 12 = -3.\n6. The fifth subtraction gives -3 - 12 = -5.\n7. The sixth subtraction gives -5 - 12 = -7.\n8. The seventh subtraction gives -7 - 12 = -9.\n9. The eighth subtraction gives -9 - 12 = -11.\n10. The ninth subtraction gives -11 - 12 = -13.\n11. The tenth subtraction gives -13 - 12 = -15.\n12. The eleventh subtraction gives -15 - 12 = -17.\n13. The thirteenth subtraction gives -17 - 12 = -19.\n14. The fifteenth subtraction gives -19 - 12 = -21.\n15. The seventeenth subtraction gives -21 - 12 = -23.\n16. The eighteenth subtraction gives -23 - 12 = -25.\n17. The nineteenth subtraction gives -25 - 12 = -27.\n18. The twentieth subtraction gives -27 - 12 = -29.\n19. The twenty-first subtraction gives -29 - 12 = -31.\n20. The twenty-second subtraction gives -31 - 12 = -33.\n21. The twenty-third subtraction gives -33 - 12 = -35.\n22. The twenty-fourth subtraction gives -35 - 12 = -37.\n23. The twenty-fifth subtraction gives -37 - 12 = -39.\n24. The twenty-sixth subtraction gives -39 - 12 = -41.\n25. The twenty-seventh subtraction gives -41 - 12 = -43.\n26. The twenty-eighth subtraction gives -43 - 12 = -45.\n27. The twenty-ninth subtraction gives -45 - 12 = -47.\n28. The thirtieth subtraction gives -47 - 12 = -49.\n29. The thirty-first subtraction gives -49 - 12 = -51.\n30. The thirty-second subtraction gives -51 - 12 = -53.\n31. The thirty-third subtraction gives -53 - 12 = -55.\n32. The thirty-fourth subtraction gives -55 - 12 = -57.\n33. The thirty-fifth subtraction gives -57 - 12 = -59.\n34. The thirty-sixth subtraction gives -59 - 12 = -61.\n35. The thirty-seventh subtraction gives -61 - 12 = -63.\n36. The thirty-eighth subtraction gives -63 - 12 = -65.\n37. The thirty-ninth subtraction gives -65 - 12 = -67.\n38. The fortieth subtraction gives -67 - 12 = -69.\n39. The forty-first subtraction gives -69 - 12 = -71.\n40. The forty-second subtraction gives -71 - 12 = -73.\n41. The forty-third subtraction gives -73 - 12 = -75.\n42. The forty-fourth subtraction gives -75 - 12 = -77.\n43. The forty-fifth subtraction gives -77 - 12 = -79.\n44. The forty-sixth subtraction gives -79 - 12 = -81.\n45. The forty-seventh subtraction gives -81 - 12 = -83.\n46. The forty-eighth subtraction gives -83 - 12 = -85.\n47. The forty-ninth subtraction gives -85 - 12 = -87.\n48. The fiftieth subtraction gives -87 - 12 = -89.\n49. The fifty-first subtraction gives -89 - 12 = -91.\n50. The fifty-second subtraction gives -91 - 12 = -93.\n51. The fifty-third subtraction gives -93 - 12 = -95.\n52. The fifty-fourth subtraction gives -95 - 12 = -97.\n53. The fifty-fifth subtraction gives -97 - 12 = -99.\n54. The fifty-sixth subtraction gives -99 - 12 = -101.\n55. The fifty-seventh subtraction gives -101 - 12 = -103.\n56. The fifty-eighth subtraction gives -103 - 12 = -105.\n57. The fifty-ninth subtraction gives -105 - 12 = -107.\n58. The sixty-first subtraction gives -107 - 12 = -109.\n59. The sixty-second subtraction gives -109 - 12 = -111.\n60. The sixty-third subtraction gives -111 - 12 = -113.\n61. The sixty-fourth subtraction gives -113 - 12 = -115.\n62. The sixty-fifth subtraction gives -115 - 12 = -117.\n63. The sixty-sixth subtraction gives -117 - 12 = -119.\n64. The sixty-seventh subtraction gives -119 - 12 = -121.\n65. The sixty-eighth subtraction gives -121 - 12 = -123.\n66. The sixty-ninth subtraction gives -123 - 12 = -125.\n67. The seventy-first subtraction gives -125 - 12 = -127.\n68. The seventy-second subtraction gives -127 - 12 = -129.\n69. The seventy-third subtraction gives -129 - 12 = -131.\n70. The seventy-fourth subtraction gives -131 - 12 = -133.\n71. The seventy-fifth subtraction gives -133 - 12 = -135.\n72. The seventy-sixth subtraction gives -135 - 12 = -137.\n73. The seventy-seventh subtraction gives -137 - 12 = -139.\n74. The seventy-eighth subtraction gives -139 - 12 = -141.\n75. The seventy-ninth subtraction gives -141 - 12 = -143.\n76. The eightieth subtraction gives -143 - 12 = -145.\n77. The eighty-first subtraction gives -145 - 12 = -147.\n78. The eighty-second subtraction gives -147 - 12 = -149.\n79. The eighty-third subtraction gives -149 - 12 = -151.\n80. The eighty-fourth subtraction gives -151 - 12 = -153.\n81. The eighty-fifth subtraction gives -153 - 12 = -155.\n82. The eighty-sixth subtraction gives -155 - 12 = -157.\n83. The eighty-seventh subtraction gives -157 - 12 = -159.\n84. The eighty-eighth subtraction gives -159 - 12 = -161.\n85. The eighty-ninth subtraction gives -161 - 12 = -163.\n86. The ninetieth subtraction gives -163 - 12 = -165.\n87. The ninety-first subtraction gives -165 - 12 = -167.\n88. The ninety-second subtraction gives -167 - 12 = -169.\n89. The ninety-third subtraction gives -169 - 12 = -171.\n90. The ninety-fourth subtraction gives -171 - 12 = -173.\n91. The ninety-fifth subtraction gives -173 - 12 = -175.\n92. The ninety-sixth subtraction gives -175 - 12 = -177.\n93. The ninety-seventh subtraction gives -177 - 12 = -179.\n94. The ninety-eighth subtraction gives -179 - 12 = -181.\n95. The ninety-ninth subtraction gives -181 - 12 = -183.\n96. The one hundredth subtraction gives -183 - 12 = -185.\n97. The one hundred and first subtraction gives -185 - 12 = -187.\n98. The one hundred and second subtraction gives -187 - 12 = -189.\n99. The one hundred and third subtraction gives -189 - 12 = -191.\n100. The one hundred and fourth subtraction gives -191 - 12 = -193.\n101. The one hundred and fifth subtraction gives -193 - 12 = -195.\n102. The one hundred and sixth subtraction gives -195 - 12 = -197.\n103. The one hundred and seventh subtraction gives -197 - 12 = -199.\n104. The one hundred and eighth subtraction gives -199 - 12 = -201.\n105. The one hundred and ninth subtraction gives -201 - 12 = -203.\n106. The one hundred and tenth subtraction gives -203 - 12 = -205.\n107. The one hundred and eleventh subtraction gives -205 - 12 = -207.\n108. The one hundred and twelfth subtraction gives -207 - 12 = -209.\n109. The one hundred and thirteenth subtraction gives -209 - 12 = -211.\n110. The one hundred and fourteenth subtraction gives -211 - 12 = -213.\n111. The one hundred and fifteenth subtraction gives -213 - 12 = -215.\n112. The one hundred and seventeenth subtraction gives -215 - 12 = -217.\n113. The one hundred and eighteenth subtraction gives -217 - 12 = -219.\n114. The one hundred and nineteenth subtraction gives -219 - 12 = -221.\n115. The one hundred and twentieth subtraction gives -221 - 12 = -223.\n116. The one hundred and twenty-first subtraction gives -223 - 12 = -225.\n117. The one hundred and twenty-second subtraction gives -225 - 12 = -227.\n118. The one hundred and twenty-third subtraction gives -227 - 12 = -229.\n119. The one hundred and twenty-fourth subtraction gives -229 - 12 = -231.\n120. The one hundred and twenty-fifth subtraction gives -231 - 12 = -233.\n121. The one hundred and twenty-sixth subtraction gives -233 - 12 = -235.\n122. The one hundred and twenty-seventh subtraction gives -235 - 12 = -237.\n123. The one hundred and twenty-eighth subtraction gives -237 - 12 = -239.\n124. The one hundred and twenty-ninth subtraction gives -239 - 12 = -241.\n125. The one hundred and thirtieth subtraction gives -241 - 12 = -243.\n126. The one hundred and thirty-first subtraction gives -24", "input_token_count": 127623, "output_token_count": 3372, "latency": 116.94029116630554}

\end{lstlisting}

\subsection{Challenge 2: Long Tool Responses}

\textbf{Verbose Responses:} For some models, the model responses are very verbose and do not follow format instructions.
For example, Qwen/QwQ-32B fails to output the precise answer and produce a lot of unsolicited text. The prompts include very specific instructions on what to output, but the models do not follow those. The example below shows Qwen/QwQ-32B being verbose even when instructed to output only the answer:

\begin{lstlisting}[basicstyle=\ttfamily, breaklines=true]
------------------------------------------------------------------
Question: Get the shortest time duration for a one-way flight from LAX to CDG. Only output the time duration in seconds without the unit and no other information.
------------------------------------------------------------------
Ground truth: 39300
------------------------------------------------------------------
Model Response:  To find the shortest time duration for a one-way flight from LAX to CDG, I'll need to look through the provided JSON data for all flight options between those airports and calculate the total time for each segment. Let me start by identifying all flight offers that start at LAX and end at CDG.

Looking at the first part of the JSON data under the key starting with "{'cabinClass': 'ECONOMY', 'sort': 'CHEAPEST', 'legs': [{'fromId': 'LAX.CITY', 'toId': 'PAR.CITY', 'date': '2024-09-18'}, ...}", there are multiple flight offers. The first flight offer has segments. The first segment is from LAX to CDG with departureTime "2024-09-18T08:30:00" and arrivalTime "2024-09-19T07:05:00". Calculating the duration: from 8:30 AM to 7:05 AM next day is 22 hours 35 minutes, which is 22*3600 + 35*60 = 81300
------------------------------------------------------------------
\end{lstlisting}

\begin{figure}[tbh]

% \vspace{-2in}
    \begin{subfigure}{0.32\textwidth}
    \includegraphics[width=\textwidth]{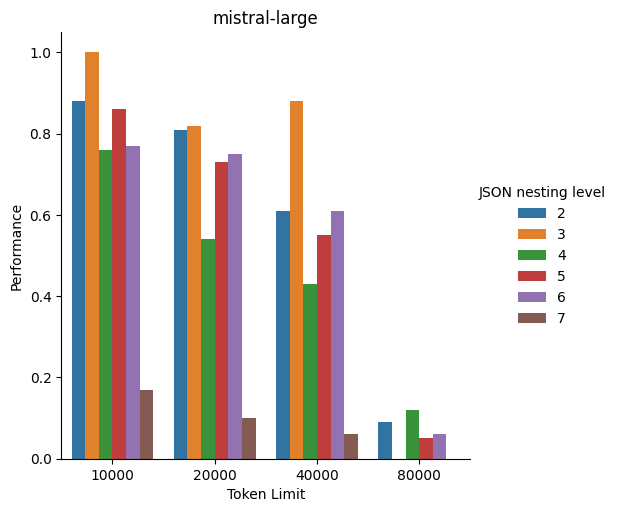}
       \caption{Mistral-Large-Instruct-2411}
    \end{subfigure}
    \begin{subfigure}{0.32\textwidth}
    \includegraphics[width=\textwidth]{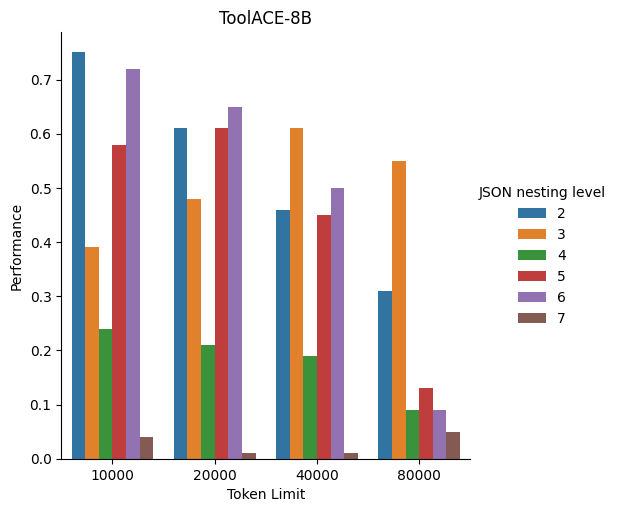}
       \caption{ToolACE-8B}
    \end{subfigure}
    \begin{subfigure}{0.32\textwidth}
    \includegraphics[width=\textwidth]{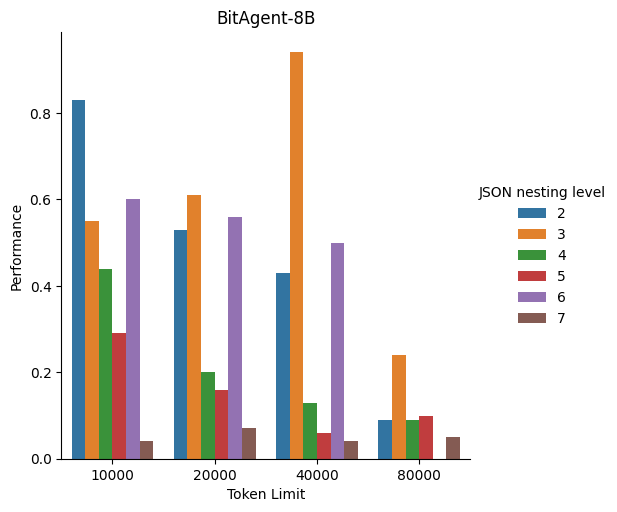}
       \caption{BitAgent-8B}
    \end{subfigure}

       \bigskip % <-- new 
       
    \begin{subfigure}{0.32\textwidth}
    \includegraphics[width=\textwidth]{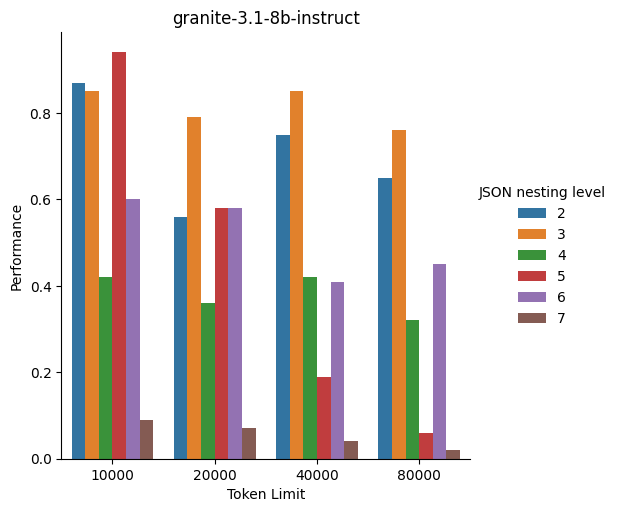}
       \caption{Granite-3.1-8b-instruct}
    \end{subfigure}
    \begin{subfigure}{0.32\textwidth}
    \includegraphics[width=\textwidth]{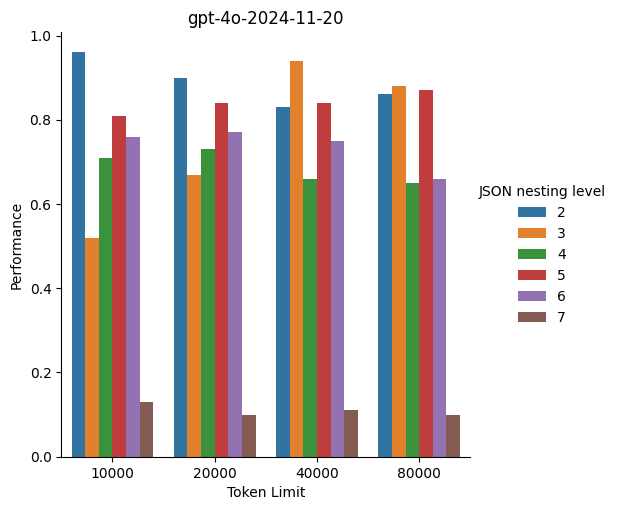}
       \caption{GPT-4o-2024-11-20}
    \end{subfigure}\hfill
    \begin{subfigure}{0.32\textwidth}
    \includegraphics[width=\textwidth]{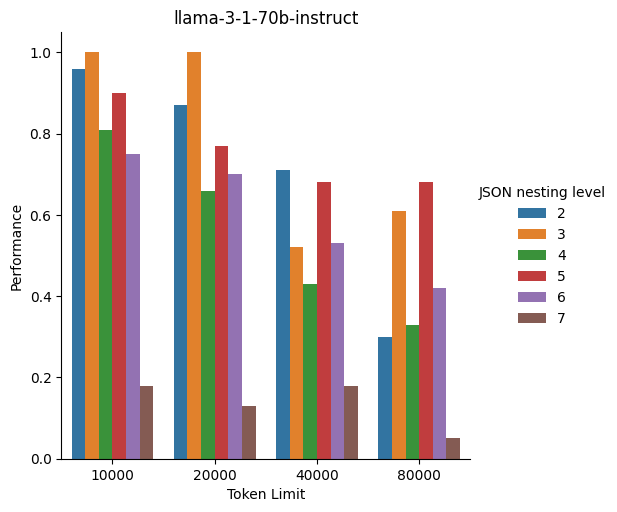}
       \caption{Llama-3.1-70B-instruct}
    \end{subfigure}

    \caption{Challenge 2: Performance based on nesting level of the answer in the JSON response. The nesting level here is calculated based on how many steps are required in the JSON traversal to get to the ground truth answer.}
    \label{fig:error_nesting}
    \end{figure}

\textbf{Nesting Level of JSON Objects:} REST API responses are complex and highly nested. We wanted to understand if the nesting level has any impact on the performance of the JSON QA task defined in Challenge 2. Figure \ref{fig:error_nesting} shows how each model performs on the question-answer pairs based on nesting level. The nesting level here is calculated based on how many steps are required in the JSON traversal to get to the ground truth answer. It can be seen that even though nesting level of 7 corresponds to low performance across models, there is no clear trend based on nesting level. This can be attributed to the fact that we are not controlling nesting levels by keeping the API responses and question-answer pairs same. So the performance is also dependent on the API responses and questions.

